\documentclass[useAMS,usenatbib]{mnras}

\usepackage{gensymb}
\usepackage{aas_macros}
\usepackage{amsmath}
\usepackage{amssymb}
\usepackage{float}
\usepackage{graphicx}
\usepackage{lscape}
\usepackage{url}
\usepackage{natbib}
\usepackage{placeins}
\usepackage{color}
\usepackage{wasysym}
\usepackage{hyperref}
\usepackage{scrextend}

\newcommand{\nhii}{\ensuremath{n_{{\mbox{\small{H}}}_{\mbox{\small{II}}}}}}
\newcommand{\nhi}{\ensuremath{n_{\mbox{\small{H}}}}}

\hypersetup{draft}
\begin{document}

\title[Photoevaporation of Exoplanet Atmospheres]
{Photoevaporative Flows From Exoplanet Atmospheres: A 3-D Radiative Hydrodynamic Parameter Study}

\author[A. Debrecht et al.]{Alex Debrecht$^{1}$\thanks{adebrech@ur.rochester.edu}, Jonathan Carroll-Nellenback$^{1}$, Adam Frank$^{1}$, 
\newauthor
John McCann$^{2}$, Ruth Murray-Clay$^{3}$, Eric G.~Blackman$^{1}$\\
$^{1}$Department of Physics and Astronomy, University of Rochester, Rochester NY 14627\\
$^{2}$Department of Physics, University of California, Santa Barbara, Santa Barbara, CA, 93106\\
$^{3}$Physics and Astronomy Department, University of California, Santa Cruz, Santa Cruz, CA 95064\\
}

\date{}

\pagerange{\pageref{firstpage}--\pageref{lastpage}}
\maketitle
\label{firstpage}

\begin{abstract}
The photoionization-driven evaporation of planetary atmospheres has emerged as a potentially fundamental process for planets on short period orbits. While 1-D studies have proven the effectiveness of stellar fluxes at altering the atmospheric mass and composition for sub-Jupiter mass planets, there remains much that is uncertain with regard to the larger-scale, multidimensional nature of such "planetary wind" flows. In this paper we use a new radiation-hydrodynamic platform to simulate atmospheric evaporative flows. Using the AstroBEAR AMR multiphysics code in a co-rotating frame centered on the planet, we model the transfer of ionizing photons into the atmosphere, the subsequent launch of the wind and the wind's large scale evolution subject to tidal and non-inertial forces. We run simulations for planets of 0.263 and 0.07 Jupiter masses and stellar fluxes of $2\times10^{13}$ and $2\times10^{14}$ photons/cm$^2$/s. Our results reveal new, potentially observable planetary wind flow patterns, including the development, in some cases, of an extended neutral tail lagging behind the planet in its orbit.
\end{abstract}

\begin{keywords}
hydrodynamics -- planet-star interactions -- planets and satellites: atmospheres
\end{keywords}

\section{Introduction}
The characterization of planetary atmospheres comprises a set of challenges at the forefront of exoplanet science. While a great deal can be learned from studying planetary atmospheres themselves, interactions between the atmosphere and the near-space environment of exoplanets is also a rich subject, promising a strong interplay between observational diagnostics and theoretical interpretation. In particular, atmospheric “blow-off”, also known as hydrodynamic escape or evaporation, may hold the key to understanding important facets of planetary evolution, from end-state masses to final atmospheric compositions (and therefore habitability).

Importantly, for some exoplanets (e.g. HD~209458b, \citet{charbonneau02,vidalmadjar03,vidalmadjar04,vidalmadjar13,ballester07,benjaffel07,benjaffel08,benjaffel10}), the observational signatures of atmospheric blow-off have become well-determined. For the hot Jupiters that have been observed to have signatures of this blow-off, from 5\% to 50\% absorption has been seen both post- and pre-transit, with some systems showing symmetric absorption (e.g. HD~209458b) and some highly asymmetric (e.g. GJ~436b, \citet{kulow14,ehrenreich15}). Interestingly, this absorption extends well into the Doppler-shifted wings of the Lyman-$\alpha$ line, with absorption out to Doppler shifts of $\pm150 \mbox{ km/s}$.

Hydrodynamic planetary winds occur when irradiation from the central star, especially in the extreme ultraviolet (EUV), heats the upper layers of the atmosphere to produce an extended envelope of gas which transitions into a wind \citep{garciamunoz07,murrayclay09}. A characteristic measure of the strength of the wind is the ratio of gravitational potential to thermal energy at the top of the atmosphere. This is usually called the hydrodynamic escape parameter $\lambda = \frac{G M_p \mu}{R_p k T_p}$, where $M_P$ and $R_p$ are the mass and radius of the planet, and $T_p$ and $\mu$ are the temperature and mean mass per particle in the atmosphere. For $\lambda >> 10$, the atmosphere is too tightly bound for a hydrodynamic wind to form. Note that weaker outflows may be produced via nonthermal processes, e.g. \citet{hunten82}. For $\lambda \sim 10$, a Parker-type thermally driven hydrodynamic wind is expected (note that $\lambda \approx 15$ for the sun with its $T \sim 10^6 \mbox{ K}$ corona). 

Physical processes that can affect the wind structure and outflow rates, such as planetary magnetic fields \citep{owen14}, time-dependent EUV flux \citep{lecavelier12}, atmospheric circulation \citep{teyssandier15}, and the interaction between stellar and planetary winds \citep{stone09,carroll16,mccann18} have been incorporated into existing simulations. Fully 3-D simulations are expensive, but a growing number of groups are carrying them out \citep{Schneiter2007, cohen11, bisikalo13, tripathi15, matsakos15, carroll16, Schneiter2017}. The work of \citet{matsakos15} was noteworthy for including a variety of processes in a 3-D MHD study that tracked the full orbital dynamics of the wind using a simplified atmospheric model.

Resolving spatial and temporal structures (asymmetry, shocks, cometary tails via radiation pressure or stellar wind interactions) in exoplanet wind simulations is of particular importance for interpreting observations. \cite{carroll16} simulated the full 3-D global dynamics of a planetary wind interacting with a stellar wind. Using the AMR code AstroBEAR \citet{carroll16} found a "2-armed" up-orbit and down-orbit pattern similar to that of \citet{matsakos15} and traced the origin of the flow to a combination of tidal and Coriolis forces. \citet{carroll16} also mapped out the observational consequences of these global flows, showing that the torus of planetary wind material was potentially observable far from the time of a transit. This conclusion was particularly relevant for WASP-12b, whose central star shows none of the expected MgII h\&k lines \citep{fossati2010a}. Their absence has been posited to be due to the presence of a relatively dense torus \citep{haswell12}. In \citet{debrecht18}, a global planetary evaporation simulation of WASP-12b was carried out using AstroBEAR and a torus of the needed density was shown to be relatively easy to form. While these global simulation approaches show promise, for both \citet{carroll16} and \citet{matsakos15} the planetary winds were not calculated self-consistently but were imposed by boundary conditions at the planet's surface.

Fully 3-D simulations that include the radiative transfer of stellar photons into the atmosphere, with its subsequent launch of the wind, are still relatively rare. In \citet{tripathi15} a self-consistent wind was driven by following the stellar flux as it was absorbed by the planet. These static mesh refinement simulations created winds whose properties were in good agreement with 1-D analytic models \citep{murrayclay09}, as well as showing features such as backflow on to the night side of the planet which had also been seen in models by \citet{Frank16}. Not included in \citet{tripathi15} were the effects of orbital motion. As shown in \citet{matsakos15} and \citet{carroll16}, the Coriolis effect strongly distorts the streamlines of the wind after it launches, forming the up- and down-orbit arms. 

Our next step in understanding planetary photoevaporative winds is to model their launch via self-consistent 3-D radiative hydrodynamic calculations, while also increasing the global scale of simulations to capture the orbital motion of the planetary wind material. In this paper we present simulations of winds launched from the atmosphere of giant planets in both long and short period orbits, with variable planetary mass and subjected to variable incident EUV flux. We follow the radiative hydrodynamic behavior out to many planetary radii and run the simulations long enough to characterize the nature of the steady state flows. We note that a similar computational approach was taken in \citet{mccann18}. While that paper focused on the interaction of planetary with stellar winds, many of our conclusions about planetary outflows are supported by both papers. The plan of the paper is as follows: in section \ref{sec:meth} we describe the numerical methods and simulations used; in section \ref{sec:result} we present the results of these simulations; in section \ref{sec:ana} we analyze the results for important consequences; and in section \ref{sec:conc} we discuss the relationship of these simulations to previous studies and propose future avenues of inquiry.

\section{Methods and Model} \label{sec:meth}

Our simulations were conducted using AstroBEAR\footnote{https://astrobear.pas.rochester.edu/} \citep{cunningham09,carroll13}, a massively parallelized adaptive mesh refinement (AMR) code that includes a variety of multiphysics solvers, such as self-gravity, heat conduction, magnetic resistivity, radiative transport, and ionization dynamics. The equations solved for these simulations are those of fluid dynamics in a rotating reference frame, with gravitational effects of both the planet and star included:
\begin{equation}
 \frac{\partial \rho}{\partial t} + \boldsymbol{\nabla} \cdot \rho \boldsymbol{v} = 0, 
 \label{eq:Eu1}
\end{equation}
\begin{equation}
 \frac{\partial \rho \boldsymbol{v}}{\partial t} + \boldsymbol{\nabla} \cdot \left ( \rho \boldsymbol{v} \otimes \boldsymbol{v} \right )= - \boldsymbol{\nabla} p - \rho \boldsymbol{\nabla} \phi + \boldsymbol{f_R},
 \label{eq:Eu2}
\end{equation}
\begin{equation}
 \frac{\partial E}{\partial t} + \boldsymbol{\nabla} \cdot ((E + p) \boldsymbol{v}) = \mathcal{G} - \mathcal{L},
 \label{eq:Eu3}
\end{equation}
where $\rho$ is the mass density, $\boldsymbol{v}$ is the fluid velocity, $p$ is the thermal pressure, $\phi$ is the gravitational potential, $\boldsymbol{f_R}$ combines the the Coriolis and centrifugal forces, so that $\boldsymbol{f_R} = \rho \left ( - 2 \boldsymbol{\Omega} \times \boldsymbol{v} - \boldsymbol{\Omega} \times \left ( \boldsymbol{\Omega} \times \boldsymbol{r} \right ) \right )$ (where $\boldsymbol{\Omega}$ is the orbital velocity), $E = p/(\gamma - 1) + \rho v^2/2$ is the combined internal and kinetic energies, and $\mathcal{G}$ and $\mathcal{L}$ are the heating and cooling rates.

The simulation also tracked the advection, photoionization, and recombination of neutral and ionized hydrogen. We use the photon-conserving update scheme from \citet{krumholz07} to solve the following equations:
\begin{equation}
 \frac{\partial \nhi}{\partial t} + \boldsymbol{\nabla} \cdot (\nhi \boldsymbol{v}) = \mathcal{R} - \mathcal{I},
 \label{eq:Eu4}
\end{equation} 
\begin{equation}
 \frac{\partial \nhii}{\partial t} + \boldsymbol{\nabla} \cdot (\nhii \boldsymbol{v}) = \mathcal{I} - \mathcal{R},
\end{equation}
where $\nhi$ is the number density of neutral hydrogen, $\nhii$ is the number density of ionized hydrogen, and $\mathcal{R}$ and $\mathcal{I}$ are the recombination and ionization rates.

\subsection{Radiation transfer} \label{sec:rad_trans}

In these simulations we model the incident stellar radiation as a planar radiation front. This is an acceptable approximation, as the spherical dilution of the radiation field is only 3\% over the simulation domain. In addition, for computational simplicity we use photons of a single frequency of 16 eV, representative of the integrated EUV flux of the quiet sun \citep{woods98}. Following \citet{tripathi15}, and as in \citet{mccann18}, we therefore apply a constant, uniform, monochromatic flux $F_0$ to the leftmost (stellar) side of the grid. The flux at a distance $x$ from the edge of the simulation is then
\begin{equation}
 F(x)=F_0e^{-\tau(x)},
\end{equation}
where $\tau(x)$ is the optical depth, given by
\begin{equation}
 \tau(x)=\int_0^x \nhi \sigma_{ph} dl.
\end{equation}
Here $\sigma_{ph} = 6.3\times10^{-18} \mbox{cm}^2$ is the cross-section for photoionization at the ionization threshold energy, $13.6 \mbox{eV}$. The photoionization rate can then be calculated:
\begin{equation}
 \mathcal{I}(x) = \sigma_{ph}\nhi F(x).
\end{equation}
As discussed in \citet{krumholz07}, this method conserves photon number. The photon flux from each previous cell is propagated, the number of absorbed photons is calculated from the optical depth of the cell, and that number is subtracted from the flux, which is then propagated forward again, so no photons are lost between cells.

We assume case-B recombination (ignoring free to ground state transitions), giving a recombination rate $\mathcal{R}$ of
\begin{equation}
 \mathcal{R} = \alpha_B(T) n_e\nhii,
\end{equation}
where the case-B recombination coefficient is $\alpha_B(T) = 2.59\times10^{-13} ({T/10^4 \mbox{K}})^{-0.7}\mbox{ cm}^3\mbox{s}^{-1}$ and $n_e$ is the number density of electrons. We also assume electrons are advected with the ionized hydrogen, so that $n_e=\nhii$. It should be noted that case-B recombination is a poor approximation throughout most of the simulation, as the winds are optically thin. However, case B is appropriate at the base of the wind, where we are most concerned with determining accurate conditions in order to obtain the correct wind solution. In addition, the difference in ionization state resulting from assuming case B has few practical consequences, as we don't perform synthetic observations in this study.

Each photon deposits energy above the ionization threshold as heat, so the photoionization heating rate is given by
\begin{equation}
 \mathcal{G} = e_\gamma \mathcal{I},
\end{equation}
with $e_\gamma = 2.4 \mbox{ eV}$ the average EUV photon energy in excess of the ionization energy. We also include the cooling from both recombination and Lyman-$\alpha$ emission due to collisional excitations:
\begin{equation}
 \mathcal{L}_{rec} = 6.11\times10^{-10} \mbox{cm}^3\mbox{s}^{-1} k_B T \left(\frac{T}{\mbox{K}}\right)^{-0.89}n_e\nhii,
\end{equation}
\begin{equation}
 \mathcal{L}_{Ly\alpha} = 7.5\times10^{-19} \mbox{erg }\mbox{cm}^3\mbox{s}^{-1} e^{-118348 \mbox{ \tiny K}/T}n_e\nhii.
\end{equation}

In order to test our radiation transfer implementation, we initialized a one-dimensional simulation at the theoretical equilibrium ionization state for a given flux and hydrogen mass density. The simulation was allowed to evolve for many crossing times, and after a small initial relaxation period was found to be extremely stable.

\subsection{Description of simulation}

The input parameters of the simulation were chosen to test the response of the planetary wind to two key parameters: planetary mass and stellar flux. Our planet, which is highly-inflated for computational reasons, has a radius of $2.146 R_{J}$ and a mass of either $0.07 M_{J}$ or $0.263 M_{J}$ (see Table \ref{tab:runs}). The planet orbits a star with a mass of $1.35 M_{\astrosun}$ at a separation of $a = 0.047 \mbox{AU}$. Table \ref{tab:runs} lists the parameters used for each simulation.

The Cartesian simulation domain ranges from $[37,-10,-10] R_p$ to $[57,10,10] R_p$, with the planet located in the center at $[47,0,0]$. We apply outflow-only extrapolating boundary conditions at all boundaries, with the initial ambient conditions applied if the extrapolated conditions would result in inflow. The simulation has a base resolution of $80^3$ and 3 levels of additional refinement, giving an effective resolution of $640^3$. The maximum resolution is forced in a sphere out to the termination radius of the hydrostatic atmosphere (defined further in Section \ref{sec:atmo}). We allow the mesh to evolve outside of the planet based on the density gradient. The planetary radius is therefore resolved by 32 cells. The stellar location $[0,0,0]$ is not included in these simulations; only the gravitational effects of the star are simulated.

Runs 1 and 2 were each performed in both the co-rotating and non-rotating frames of reference. The planet temperatures were adjusted based on the planet mass in order to maintain reasonable scale heights near the surface of the planet. It was shown in \citet{murrayclay09} (Appendix A) that the planet's surface temperature had an insignificant effect on the wind structure for any temperature below the wind temperature at the base, generally around $10^4 \mbox{K}$.

The surface densities of the planet were adjusted so that the surface with optical depth one in the unperturbed planet was near the nominal planet surface, $R_p$. This was initially calculated for the low-flux case, then scaled by $\sqrt{F_{0,2}/F_{0,1}}$ for the high-flux case, based on the assumption that we are in the recombination-limited regime of \citet{murrayclay09}. Although this is shown to be false in section \ref{sec:regime}, it provides an acceptable estimate for the required change in density. Our simulations were run for 5.24 days (1.49 orbits), after which the outflows have all reached a steady state, defined as a stable ionization front and wind morphology.

\setcounter{table}{0}

\begin{table*}
\centering

 \caption{Run parameters}
 \label{tab:runs}
 \begin{tabular}{l|c|c|c|c|c|}

  \hline

  & & Run 1 & Run 2 & Run 3 & Run 4 \\

  \hline
  
  Planet Radius & $R_p$ & \multicolumn{4}{c}{$2.146 R_{J}$} \\
  Stellar Mass & $M_\star$ & \multicolumn{4}{c}{$1.35 M_{\astrosun}$} \\
  Orbital Separation & $a$ & \multicolumn{4}{c}{$0.047$ AU} \\
  Orbital Period & $P$ & \multicolumn{4}{c}{$3.525$ days} \\
  Orbital Velocity & $\Omega$ & \multicolumn{4}{c}{$1.78$ rad/day} \\
  Polytropic Index & $\gamma$ & \multicolumn{4}{c}{$\frac53$} \\
  Planet Mass & $M_p$ & $0.07 M_{J}$ & $0.263 M_{J}$ & $0.07 M_{J}$ & $0.263 M_{J}$ \\
  Planet Temperature & $T_p$ & $1\times10^3$ K & $3\times10^3$ K & $1\times10^3$ K & $3\times10^3$ K\\
  Stellar Ionizing Flux & $F_0$ & $2\times10^{13} \mbox{ phot cm}^{-2}\mbox{s}^{-1}$ & $2\times10^{13} \mbox{ phot cm}^{-2}\mbox{s}^{-1}$ & $2\times10^{14} \mbox{ phot cm}^{-2}\mbox{s}^{-1}$ & $2\times10^{14} \mbox{ phot cm}^{-2}\mbox{s}^{-1}$\\
  Planet Surface Density & $\rho_p$ & $1.324\times10^{-16} \mbox{ g cm}^{-3}$ & $1.324\times10^{-16} \mbox{g cm}^{-3}$ & $5.05\times10^{-16} \mbox{g cm}^{-3}$ & $5.05\times10^{-16} \mbox{g cm}^{-3}$\\
  
  \hline

 \end{tabular}
\end{table*}

\subsection{Planet atmosphere model} \label{sec:atmo}

In these simulations, we have modeled the planet as a sphere of hydrogen held in hydrostatic equilibrium so that it satisfies
\begin{equation}
 \frac{dP}{dr} = -\frac{G M_p \rho}{r^2},
\end{equation}
giving a density profile of
\begin{equation}
 \rho_{atm}(r) = \left [\frac{R_0 R_p}{R_0 - R_p} \left (\frac1r -\frac1{R_0}\right)\right]^{\frac1{\gamma-1}}.
\end{equation}
In the equations above
\begin{equation}
 R_0 = \frac{R_p}{1 - (\gamma c_{s,p}^2 R_p)/([\gamma - 1] G M_p)}
\end{equation}
is the radius at which the atmospheric profile goes to zero (physically, the atmosphere ends), with $c_{s,p} = \sqrt{\frac{k_B T_p}{m_H}}$ the speed of sound at $R_p$. Since we have taken an adiabatic index of $\gamma = \frac53$, the hydrostatic pressure profile is given by
\begin{equation}
 P_{atm}(r) = C \rho^\gamma,
\end{equation}
with C a constant set by the pressure at the planet surface:
\begin{equation}
 P(R_p) = \frac{\rho_p k_B T_p}{m_H},
\end{equation}
for the ideal gas case. This gives $C=\rho_p^{1-\gamma} c_{s,p}^2$. 

In order to prevent the singularity at $r = 0$, we cut off the planet profile at $r = R_{mask} = 0.2 R_p$. Interior to $r = R_{ib} = 0.35 R_p$, 5 grid cells outside $R_{mask}$, we reset the planet profile at every time step in order to replenish the supply of planet material blown out by the wind. Finally, at the outer boundary $r = R_{ob} = 1.35 R_p$, we cut the planet profile off on the outside edge and set up a static ambient with pressure matched to the final value of the planet profile, completely specifying the initial conditions of the simulation:
\begin{equation}
 \rho(r) = 
 \begin{cases}
  \rho_{atm}(R_{mask}), & r < R_{mask} \\
  \rho_{atm}(r), & R_{mask} \leq r \leq R_{ob} \\
  \rho_{atm}(R_{ob}) \cdot 10^{-4}, & r > R_{ob}
 \end{cases},
\end{equation}
with corresponding pressures of
\begin{equation}
 P(r) = 
 \begin{cases}
  P_{atm}(R_{mask}), & r < R_{mask} \\
  P_{atm}(r), & R_{mask} \leq r \leq R_{ob} \\
  P_{atm}(R_{ob}), & r > R_{ob}
 \end{cases}.
\end{equation}

\section{Results} \label{sec:result}

In what follows we present the results of the simulations in terms of their steady state flow characteristics. We focus on hydrodynamic flow patterns and ionization conditions within the flows.

\subsection{Long-period planets: Low Mass, Low Flux case}

All of our simulations are centered on the planet and carried out in the planet's orbiting frame of reference. The first set of two simulations we discuss are for cases where no non-inertial or tidal forces are applied. Although the primary purpose of this paper is to study the behavior of short period giant planets (for which these forces will be significant), the non-rotating cases allow us to make contact with previous work \citep{tripathi15}, as well as similar cases in \citet{mccann18}, and establish a baseline for the subsequent short period models. In addition, these simulations may be relevant to giant planets orbiting more luminous O-, B-, and potentially A-type stars at orbital radii comparable with that of Jupiter in our solar system \citep{sternberg03}. 

We begin with the long period version of Run 1, the low flux, low mass planet, shown in figure \ref{fig:rxt_run1norot}. The leftmost panel of this figure shows the density (hue, logarithmically scaled), in $\mbox{g cm}^{-3}$, and velocity field (quivers) for this simulation. In addition, in this panel we show the $\tau = 1$ surface (black contour), Mach surface (magenta contour), and nominal planetary radius $R_p$ (green contour). Because tidal and non-inertial forces are not present, the wind is symmetric for rotations about the x-axis; therefore, we show only a slice of the x-z plane cutting through the planet and looking up-orbit.

The $\tau = 1$ surface can be considered the surface of the planet from which the wind is launched. At this point, most of the ionizing photons have been absorbed. Below the $\tau = 1$ surface the planetary hydrogen density and the corresponding optical depth quickly increase such that by $R = 0.5 R_p$, 99.9\% of the incident radiation has been absorbed. \citet{murrayclay09} found that the wind solution is insensitive to conditions below the $\tau = 1$ surface. Although the details of the flow below $\tau = 1$ are not expected to accurately model conditions of a real giant planet, this region still plays a role in the simulation by providing a flow of neutral material to larger radii that is subsequently ionized and continually supplies the wind.

The $\tau = 1$ contour also denotes the extent of shadowing by the planet. In the non-rotating cases, this shadow extends uniformly in the $-x$ direction. This results in material launched from the planet's night side remaining neutral. Much of this flow also remains subsonic. By following movies\footnote{To view movies of the simulations, see the AstroBEAR YouTube page (https://www.youtube.com/user/URAstroBEAR)} of the simulations we can see that the subsonic material in the planetary shadow eventually passes off the grid, although the Mach surface in figure \ref{fig:rxt_run1norot} has yet to reach that point.

On the day side of the planet the wind follows nearly radial streamlines as expected. The Mach surface is nearly spherical on the day side and is close to the wind launching radius. Moving from the day to the night side of the planet, the sonic surface becomes aspherical due to pressure effects. Because the wind is stronger on the day side, pressure gradients drive lateral flows from one hemisphere to the other, as seen in figure 2 of \citet{tripathi15} as well as figure 3 of \citet{carroll16}. These lateral flows influence the evolution of the Mach surface. Initially the Mach surface is coincident with the $\tau=1$ surface in the planet's shadow. In time, however, the day-side wind sweeping around the planet collides with the night-side wind material, compressing it to a cylinder of radius less than $R_p$.

This effect can also be seen in the center panel of figure \ref{fig:rxt_run1norot}, which shows the neutral hydrogen fraction for the same slice through the planet. Here we see that the subsonic tail is almost entirely neutral due to the planet's ionization shadow. A linear scaling was chosen to highlight the compression of the tail material. Also seen in the center panel is the fact that the rest of the planetary wind (outside the shadow) carries a small fraction of neutrals with it. The neutral fraction $\nhi/n_H$ is of order $10^{-3}$ in the bulk of the wind. This is more obvious in the logarithmically scaled plots used for figures \ref{fig:rxt_run1} through \ref{fig:flow_texture_run4}.

The final panel in figure \ref{fig:rxt_run1norot} shows the temperature, in $\mbox{K}$, on a logarithmic scale. Because the wind is mostly transparent to ionizing radiation outside of the planet's atmosphere, the temperature of the wind is determined hydrodynamically, beginning at $3500 \mbox{ K}$ at the base of the wind and cooling primarily by expansion (though radiative and recombination cooling are still present).

The temperature map also highlights the interior portions of the planet which remain at the initial conditions. The temperature transition from the upper atmosphere to the base of the wind is sharply delineated, as would be expected from the previous discussion of the optical depth. Although the initial planet profile is maintained at a greater radius on the night side, there are some atmospheric dynamics present there as well.

We now consider the flow pattern in the model via figure \ref{fig:flow_texture_run1norot}, which was created by convolving random noise integrated along the streamlines of the velocity field with a color plot of the density. In this figure, hue shows density (as in the first panel of figure \ref{fig:rxt_run1norot}) and the texture of the plot represents flow streamlines. The nearly spherical symmetry of the flow on the day side of the planet is conspicuous. The slower, confined subsonic flow in the tail is also clearly delineated. The backflow towards the night side seen in previous studies is highlighted by this plot as well, particularly on the night side close to the planet.

\subsection{Long-period planets: High Mass, Low Flux Case}

Before we consider the results of the high mass case we note that the momentum flux of the wind is its ram pressure, which is determined in our case by the escape velocity $v_{esc}$ and the density at the base of the wind, with
\begin{equation}
 v_{esc} = \sqrt{\frac{2 G M_p}{R_p}}.
\end{equation}
In a more general sense, the outflow velocity is not necessarily set by only by the escape velocity at the base of the wind, but also by the distance it takes for the flow to accelerate to the sound speed. However, in our cases, the sonic radius and the base of the wind are located within a small fraction of a planetary radius of each other, which allows us to neglect this factor when determining the relative ram pressures of our outflows.

The wind material must be given an energy sufficient to achieve this velocity in order to be liberated from the planet's gravity. Since the (nominal) planet radius $R_p$ is the same for all of our simulations, only the planet mass affects the velocity with which the winds are launched. The escape velocity for Runs 1 and 3 is $10.9 \mbox{ km s}^{-1}$, while the escape velocity for the higher-mass Runs 2 and 4 is $21.1 \mbox{ km s}^{-1}$. This difference in the wind speed will bear directly on the models, since it affects the ram pressure of the wind.

The other factor in the ram pressure is the density of the wind. Although the density is diluted as the wind expands (mostly) spherically away from the planet, the relative density at particular radii is set by the density at the base of the wind. The parameters at the base of the wind are given in table \ref{tab:wind_prop}. The low-flux runs have densities within 20\% of each other, and the high-flux runs have densities within 10\% of each other; however, the low-flux runs differ from the high-flux runs by a factor of 4.5. Note that this is greater than the increase in density between the two sets of runs (see table \ref{tab:runs}), because the base of the wind in the high flux cases is deeper inside the planet and launching from a higher ionization fraction. While this could explain a dichotomy in the behavior of the low-flux and high-flux runs, only the increased planetary mass could produce a significant difference between the two low-flux runs.

\begin{figure*}
\centering
\includegraphics[width=\textwidth]{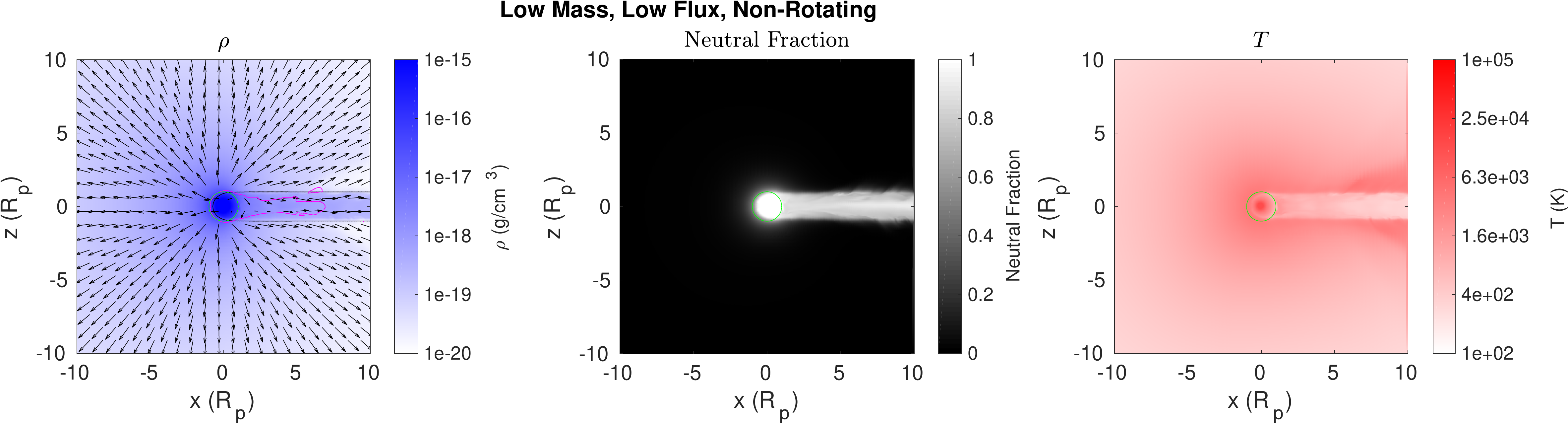}
\caption{Density (left), neutral fraction (center), and temperature (right) for Run 1 in the non-rotating case, standing in the orbital plane and looking in the up-orbit direction. The quivers describe the velocity field, and the contours are of the Mach surface (magenta), the $\tau = 1$ surface (black), and $R_p$ (green). The extended neutral tail is due to the shadow of the planet.}
\label{fig:rxt_run1norot}
\end{figure*}

\begin{figure}
\centering
\includegraphics[width=\columnwidth]{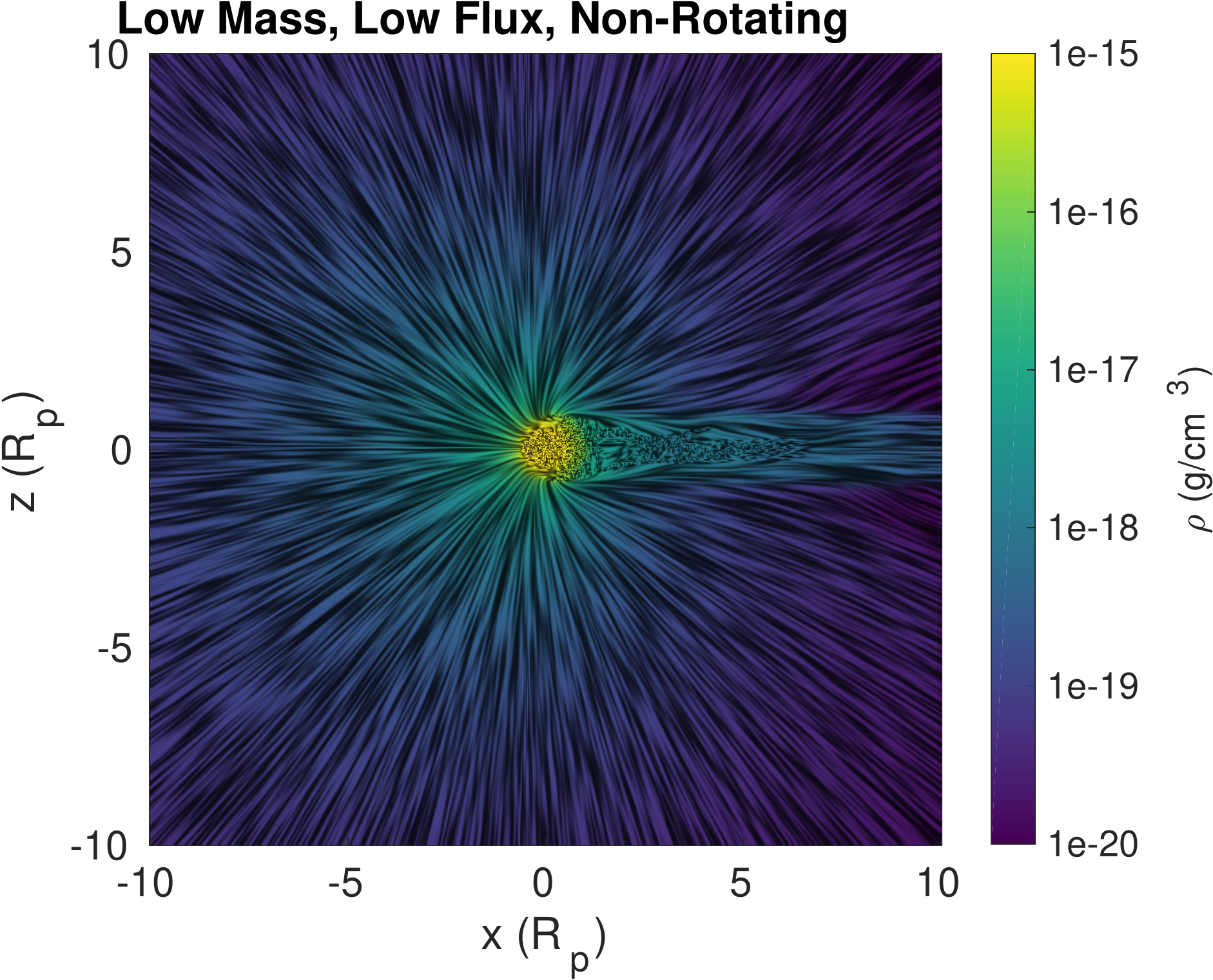}
\caption{Flow-texture plot of Run 1 in the non-rotating case, showing density (hue) and local streamline orientation (texture). The outflow from the planet is nearly spherically symmetric due to the lack of rotational forces, with a small flow around the planet from the day side to the night side.}
\label{fig:flow_texture_run1norot}
\end{figure}

\begin{figure*}
\centering
\includegraphics[width=\textwidth]{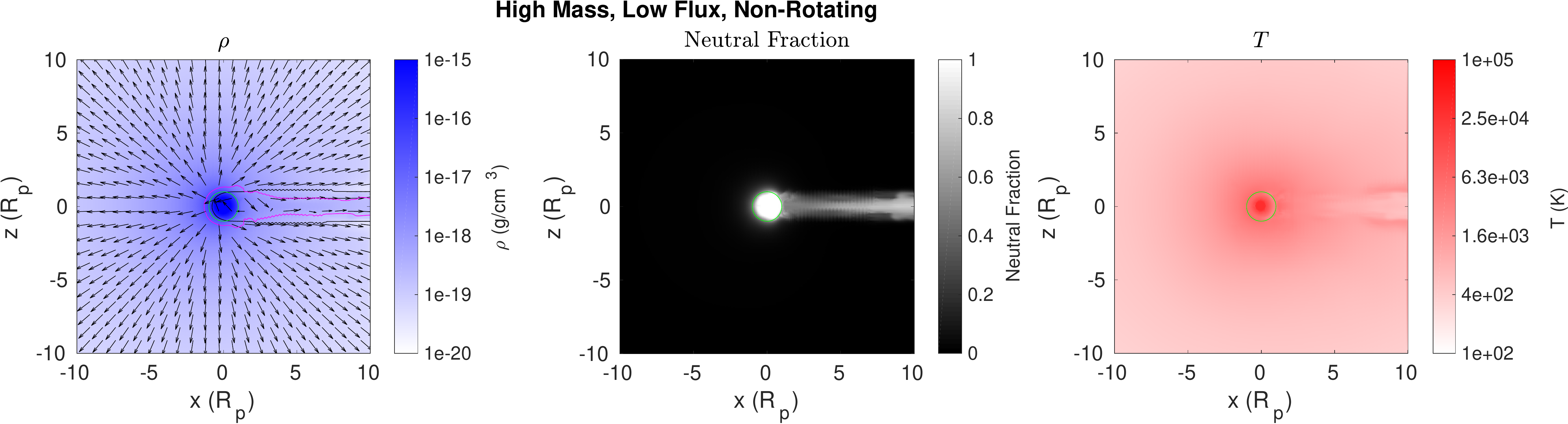}
\caption{Density (left), neutral fraction (center), and temperature (right) for Run 2 in the non-rotating case, standing in the orbital plane and looking in the up-orbit direction. The quivers describe the velocity field, and the contours are of the Mach surface (magenta), the $\tau = 1$ surface (black), and $R_p$ (green). The extended neutral tail is again due to the shadow of the planet, but in this case is diluted and compressed by the stronger planetary wind.}
\label{fig:rxt_run2norot}
\end{figure*}

We now consider the details of the long period Run 2 simulation. Figure \ref{fig:rxt_run2norot} is the same as figure \ref{fig:rxt_run1norot} for the higher-mass planet. Although the large-scale behavior is very similar, still showing a nearly spherically symmetric outflow, there are differences (particularly in the lateral flow around the planet) that will become more significant in the rotating cases. First note that although the ionization front is in a similar location, the Mach surface is located significantly outside $R_p$. In addition, while the sonic surface is still compressed by the lateral flows far from the planet, it has a width comparable to $R_p$ near the planet.

The neutral density (center panel) highlights another significant change due to the stronger wind blowing around the planet. Here we see stronger compression of the neutral tail than found previously, which is almost entirely due to the lateral flows from the day side of the planet. Finally, due to the higher planet surface temperature the ionization front cannot be seen as clearly in the rightmost panel.

\begin{figure}
\centering
\includegraphics[width=\columnwidth]{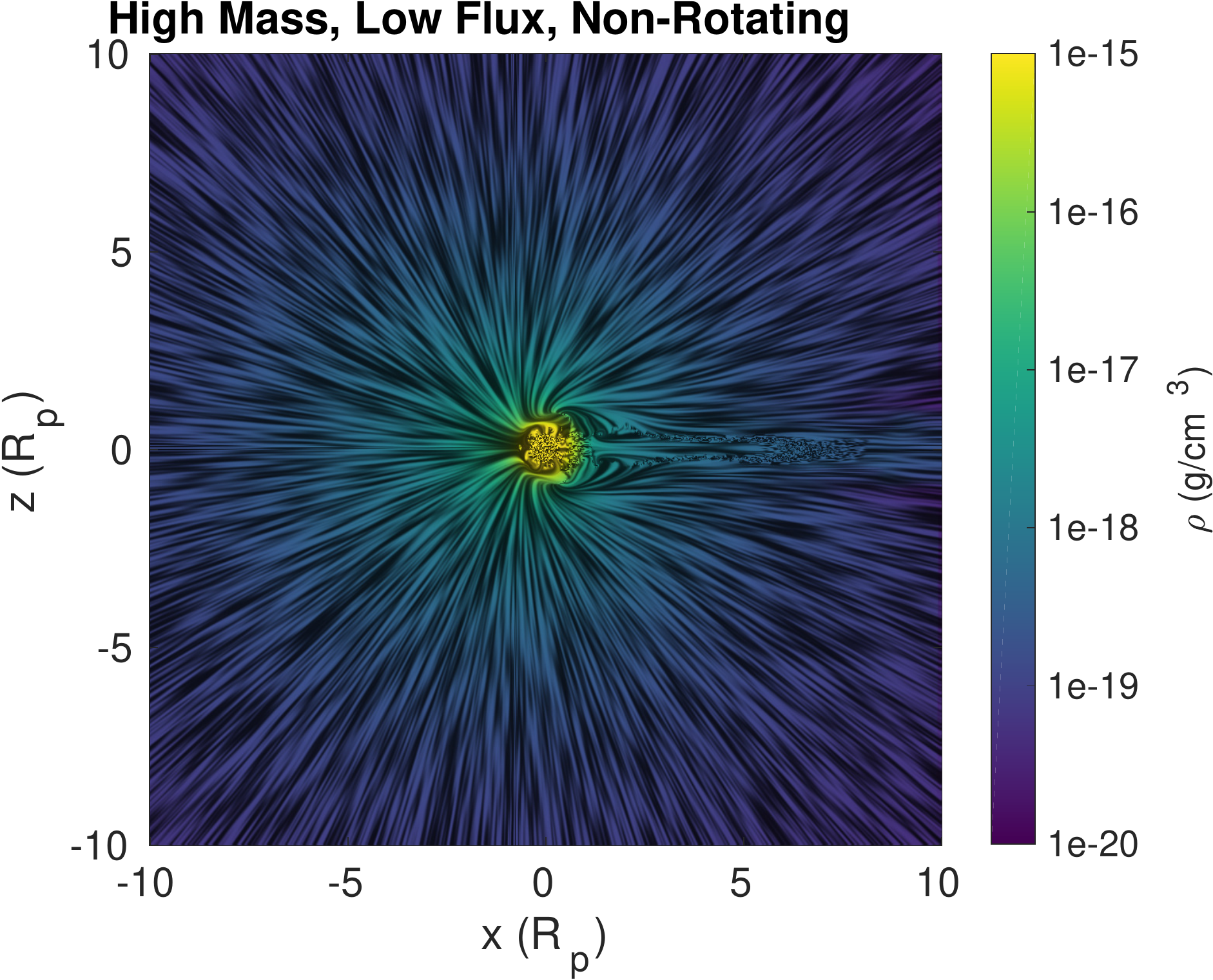}
\caption{Flow-texture plot of Run 2 in the non-rotating case, showing density (hue) and local streamline orientation (texture). The outflow is again nearly spherical, but the flow around the planet can be seen much more strongly than in Run 1.}
\label{fig:flow_texture_run2norot}
\end{figure}

Figure \ref{fig:flow_texture_run2norot} shows the flow streamlines for the higher-mass planet. This figure serves to highlight the significant flow around the planet from the day side to the night side, which alters the characteristics of the neutral tail as discussed with figure \ref{fig:rxt_run2norot}.

In general, the two long period models establish the code's ability to accurately capture the details of the launching of the wind via ionizing stellar flux as well as the resulting larger-scale flow out to many planetary radii.

\begin{table}
 \begin{minipage}{80mm}

 \caption{Atmospheric properties at base of wind}
 \label{tab:wind_prop}
 \begin{tabular}{c|c|c|c|c}

  Simulation & $T$ & $\rho$ & $R$ & $X$ \\
  & K & $\mbox{ g cm}^{-3}$ & $R_p$ & \\

  \hline

  Low Mass, Low Flux & 3300 & $1.31 \times 10^{-16}$ & 0.80 & 0.24 \\
  High Mass, Low Flux & 5302 & $1.62 \times 10^{-16}$ & 0.86 & 0.24 \\
  Low Mass, High Flux & 5312 & $5.80 \times 10^{-16}$ & 0.61 & 0.54 \\
  High Mass, Low Flux & 6856 & $5.30 \times 10^{-16}$ & 0.80 & 0.55 \\

  \hline

 \end{tabular}
 \end{minipage}
\end{table}

\subsection{Short-period planets: Low Mass Low Flux Case}

We now consider the models which comprise the main focus of this work. In \citet{carroll16} the redirection of the planetary wind into a torus-like configuration comprised of up-orbit and down-orbit arms was investigated. In that study it was found that the formation of the torus could be understood, to first order, via Coriolis forces. This was expressed via a "Coriolis length" for wind flows, given by
\begin{equation}
 \frac{R_{\Omega}}{a} = \sqrt{\frac{aq}{\lambda R_p \left(q+1 \right)}},
\end{equation}
where $a$ is the orbital radius and $q = M_p/M_*$. For small $q$ and $a$ it can be seen that the ratio $ \frac{R_{\Omega}}{a} < 1$ and the wind streamlines are turned $90\degree$ by the Coriolis force on length scales smaller than the orbit. In the \citet{carroll16} study the planetary outflow was driven from preset temperature and density conditions at a preset radius that was intended to represent the outer edge of the planet (its exobase). The day side of the planet was set to a temperature $T(\theta)=T_{p} \max \left [ 0.01, \cos (\theta) \right ]$ where $\theta$ is the angle of incidence of the light from the star (and the angular distance from the sub-stellar point). The night side was kept at $.01 T_p = 100 \mbox{K}$ \citep{stone09}. We now explore the flow patterns produced for short period planets (strong Coriolis and tidal forces) when the wind is generated self-consistently via radiation deposition in the atmosphere. 

\begin{figure*}
\centering
\includegraphics[width=\textwidth]{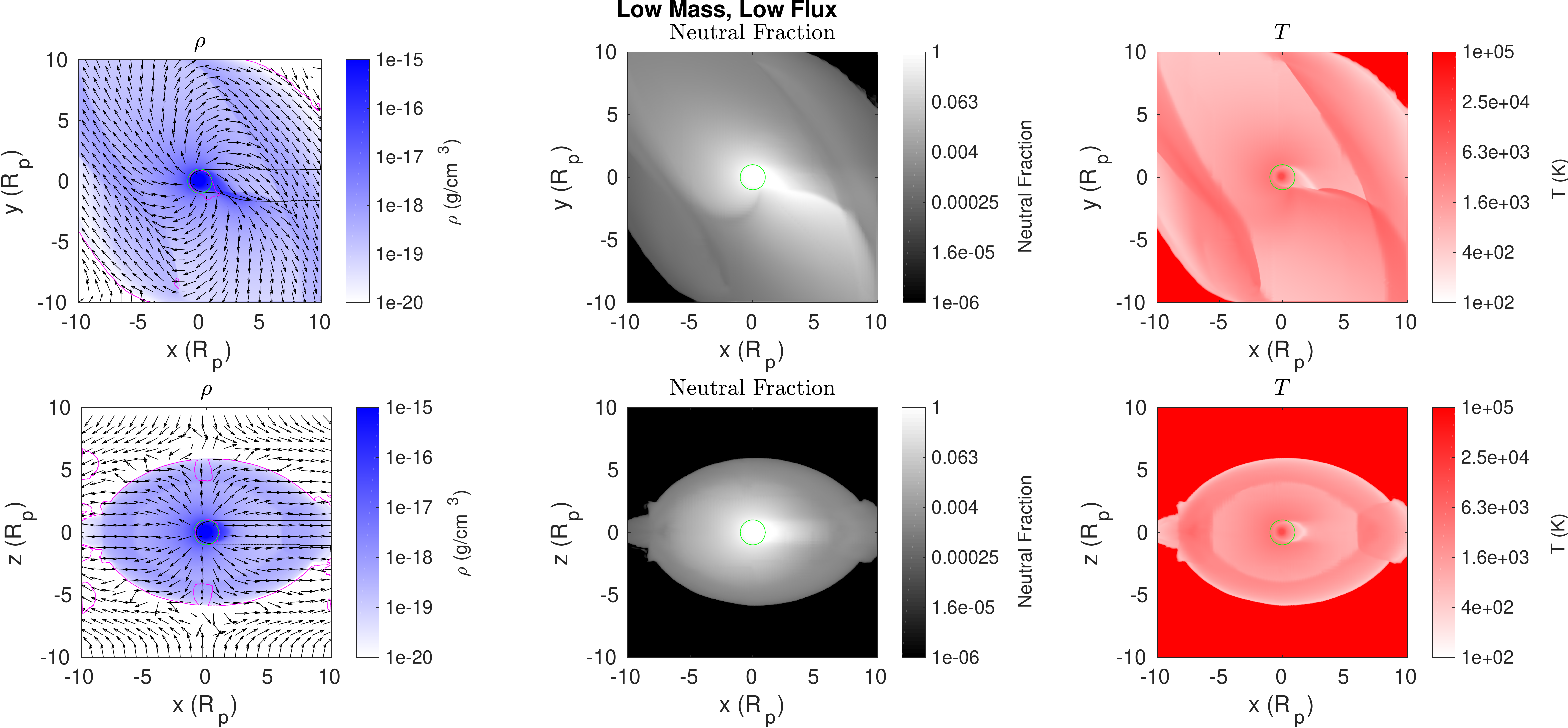}
\caption{Density (left), neutral fraction (center), and temperature (right) for Run 1 in the short period case, with a view down onto the orbital plane (top row) and standing in the orbital plane looking in the up-orbit direction (bottom row). The quivers describe the velocity field, and the contours are of the Mach surface (magenta), the $\tau = 1$ surface (black), and $R_p$ (green). There is an extended thin supersonic tail of neutral material leaving the night side of the planet, and complex oblique shock structures along the tube of the wind, which is confined in the radial and perpendicular directions.}
\label{fig:rxt_run1}
\end{figure*}

We first consider the low mass, low flux, short orbit case. Figure \ref{fig:rxt_run1} again shows the density (left), neutral fraction (center, now logarithmically scaled), and temperature (right) with contours of Mach number (magenta), optical depth (black), and planet radius (green). The simulation is again carried out in the co-rotating frame, but because we now have a rapidly orbiting planet, we show a 2-D slice looking down on the orbital plane (top row) and a side view (bottom row).

As in \citet{carroll16} and \citet{matsakos15}, we again see the formation of up- and down-orbit wind trajectories (top row). The wind flow no longer fills the whole computational domain. Rather, it is confined to a torus with a quasi-cylindrical cross section (bottom row). Note that the wind is bounded in all directions with shocks at the interface with the original ambient medium, as well as additional shocks in the interior of the flow. While the structure of the wind has changed significantly from the non-rotating cases, one important similarity can be seen on the day side of the planet in the density panel (top left). Here, the ionization front and Mach surface are nearly identical to those found in the non-rotating case, suggesting, as one would expect, that these are set only by the incident flux and the planet parameters. 

In addition to the changes in global flow structure, there is no longer a subsonic tail present in this model. A small portion of the material flowing from the night side that has been turned down-orbit does remain subsonic near the planet, which creates shocks with gas from the up-orbit arm that has been turned completely around the planet and now flows in the $-y$ direction. This interior shock creates a high-density region which can also be seen in the $\tau = 1$ surface. Rather than forming directly behind the planet, the shadow is now extended in the $-y$ direction by the tail on the down-orbit side of the planet.

The neutral fraction $\nhi/n_H$ (top center panel) shows that the tail remains neutral with an extent defined by the ionization timescale and wind velocities via advection. In this model, the neutral tail extends a significant distance from the planet. There is a similarly-shaped feature mirrored across the $y$ axis, where ionized material originating from the down-orbit side of the planet turns in the up-orbit direction. Finally, note that, as in the long period simulations, the bulk of the wind far from the ionization shadow maintains a small neutral fraction of order $10^{-2}$.

\begin{figure*}
\centering
\includegraphics[width=\textwidth]{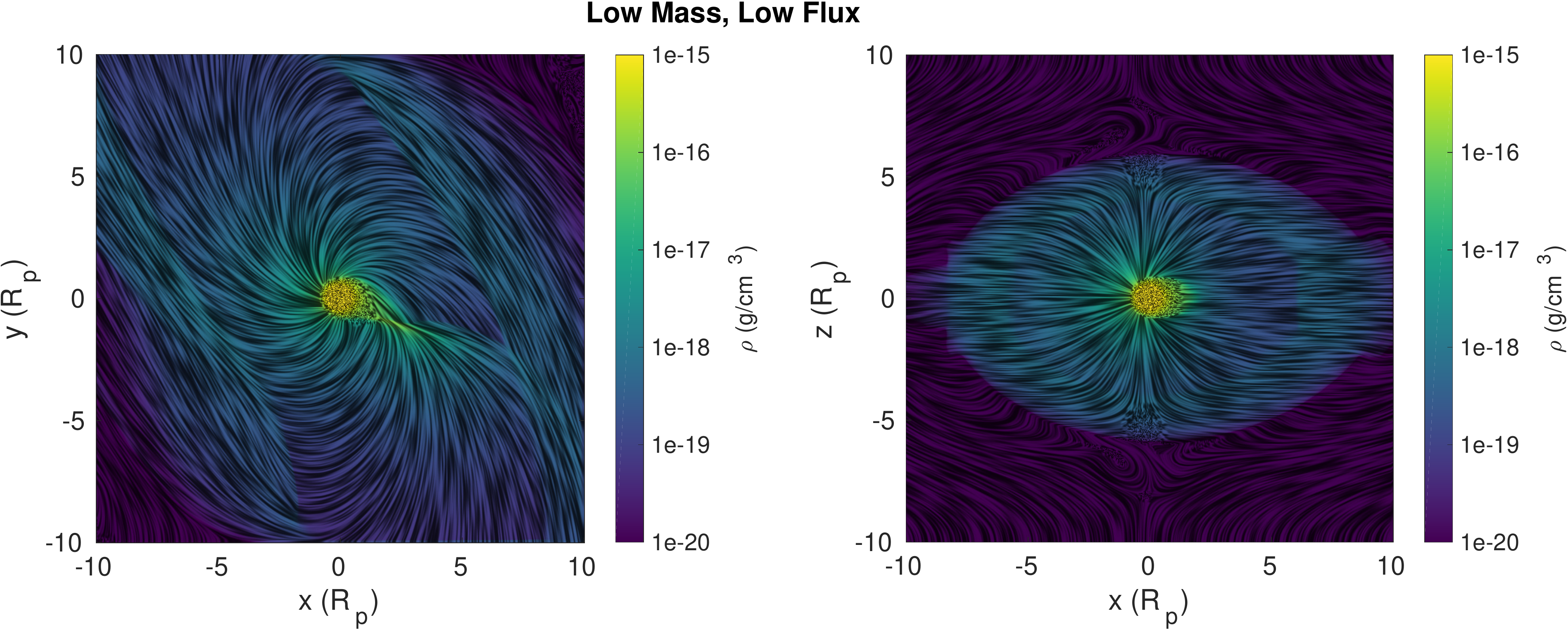}
\caption{Flow-texture plot of Run 1 in the short period case, showing density (hue) and local streamline orientation (texture), looking down onto the orbital plane (left) and standing in the orbital plane and looking up-orbit (right). The redirection of the flow at the shocks is apparent, as well as the azimuthal symmetry of the streamlines out to $\sim4 R_p$.}
\label{fig:flow_texture_run1}
\end{figure*}

The temperature (top right panel) highlights the internal shock structures present in the wind. There are three primary internal shocks of interest. The first appears in the left-most boundary of the flow. This shock is created by material leaving the planet at latitudes closer to the sub-stellar point colliding with material leaving closer to the $-y$ terminator and nightside of the planet (see also \ref{fig:flow_texture_run1}). The second shock occurs near the right edge of the flow and forms from material leaving latitudes near the sub-stellar point colliding with material leaving the closer to the $+y$ terminator and night side. Finally, we see material leaving the planet near the $+y$ terminator colliding with material leaving the night side, which forms the shock structure at the supersonic neutral tail.

The density in the $x-z$ plane (bottom left panel) shows that the rotational forces creating the up-orbit and down-orbit arms of the wind also confine it to a torus surrounding the planet. Oblique internal shocks can be seen in the corresponding temperature plot bounding the rim of the torus. Note that the $\tau = 1$ surface is once again defined by the extent of the planet in this plane. The edges of the torus in the $x-z$ plane also show some corrugations, which may be due to instabilities at the interface with the ambient medium.

The neutral fraction in the $x-z$ plane (bottom center panel) shows that the neutral tail extends through the ionization shadow of the planet. The fact that the tail does not reach the torus edge in this cut is consistent with its being driven down-orbit via the shocks visible in the cut through the orbital plane.

Figure \ref{fig:flow_texture_run1} shows the flow texture of the low flux, low mass planet run from the top view (left) and side view (right). The redirection of the flow at the shock structures is most apparent in the top view, with the origins of the shocked material simple to trace as described previously. Note the similarity of the streamlines in this plane to the semi-analytic models of figure 6 in \citet{carroll16}. In those calculations \citet{carroll16} were able to predict the wind streamlines from ballistic trajectories of wind parcels launched from the planet's surface and subject to Coriolis and tidal forces (in particular see the $\Xi_p = 0.2$ and $\tau = 1$ models as defined in that paper).

The side view shown in \ref{fig:flow_texture_run1} shows that the wind flow near the planet has a symmetric structure and that streamlines encounter the internal shocks close to the wind-ambient medium interface. The stretching of the wind due to stellar tidal forces is also apparent in the right panel. 

\subsection{Short-period planets: High Mass, Low Flux Case}

In figure \ref{fig:rxt_run2} we see the effect of raising the planet's mass while keeping the ionizing flux constant. As seen in the density (top left), significant portions of the up-orbit and down-orbit arms of the flow are now subsonic, in contrast with Run 1. Additionally, the Coriolis force confines the wind much more strongly in this case, with the boundaries of the outflow no longer extending to the edge of the grid in the $x$ direction. As in the non-rotating case, the Mach surface is located farther from the base of the wind here. Due to the lack of a strong shock enhancing the neutral density on the night side of the planet, the $\tau=1$ contour no longer protrudes significantly beyond the shadow of the planet.

The neutral fraction (top center panel) highlights one consequence of the planetary wind's stronger confinement to the up/down orbit torus. Here, the neutral tail no longer shows an extended arc behind and down-orbit from the planet. As in the non-rotating, high mass planet case (fig. \ref{fig:rxt_run2norot}), this model has a stronger wind in terms of its ram pressure $P_{ram} = \rho_w v_w^2$ which, in this case, sweeps around the planet and truncates the neutral tail. This is also shown prominently in the temperature map (top right panel), where the temperature jump seen in the low planet mass case is absent. In addition, the oblique shocks are notably weaker (and in the case of the down-orbit shocks, completely absent) here.

The slices in the $x-z$ plane (bottom set of panels) again show the strong confinement of the wind. From this perspective, we can see the torus of wind material has a smaller cross section and none of the wind leaves the grid in this plane. The density map in this plane (bottom left panel) again shows the Mach surface being held farther from the planet than in Run 1. The bottom center panel also shows a faint increase in the neutral fraction in the shadow of the planet due to wind material recombining as it passes through the shadow.

Figure \ref{fig:flow_texture_run2} shows the flow texture map for the higher-mass planet. Comparing figure \ref{fig:flow_texture_run1}, we see that the wind has a smaller turning radius. The cause of the down-orbit subsonic flow is also revealed to be material leaving in the up-orbit direction and being turned sharply down-orbit to impact material launching from the night side. The side view (right panel) also provides evidence for the larger effect of the rotational forces, with only a very small portion of the outflow appearing azimuthally symmetric when compared to figure \ref{fig:flow_texture_run2norot}.

\subsection{Short-period planets: Low Mass High Flux Case}
We now consider simulations for planets exposed to an incident flux that is an order of magnitude higher than the previous models.

Figure \ref{fig:rxt_run3} includes the same 6 panels shown previously, but this time for the low planet mass, high stellar flux case (Run 3). The overall flow pattern is similar to that seen in the low planet mass, low stellar flux case (Run 1) with a few key differences. 

In the density panel (top left) we see that the Mach surface is very close to the base of the wind, as in Run 1. We also again see a high-density neutral tail present on the night side. The shock formed from day side material sweeping around to the night side of the planet again creates a protrusion of the $\tau=1$ surface down-orbit, beyond the shadow of the planet. The total extent of the wind flow (up-orbit and down-orbit arms) is similar to Run 1 (and again larger than in Run 2). The strong oblique internal shocks (most apparent in the temperature panel) are also present as in Run 1.

One of the key differences between this high flux case and the previous low flux version can be seen in the neutral density (top center panel). In the high flux case we see a thicker neutral tail due to the the increased planetary mass flux which occurs thanks to the higher radiative flux in Run 3. From Table \ref{tab:mdot} we see that ${\dot M}_{Run 3} \sim 10 {\dot M}_{Run 1}$. In addition, the increase in neutral fraction across the shadow of the planet can be attributed to the lower equilibrium ionization state in the bulk of the wind for the high-flux runs, which leads to greater recombination in the tail (due to increased electron density) and increased contrast between the highly-ionized wind material and the recombined material in the planet's shadow.

The temperature map (top right panel) also highlights changes in the structure of the tail. Run 3 shows two shocks in the tail, one when the material turned from the up-orbit outflow shocks with the tail material, and another when the tail material shocks with the remaining night-side and down-orbit outflow (these flows are more easily seen in Fig \ref{fig:flow_texture_run3}). In addition, the temperature map highlights again the discontinuity between the base of the wind and the undisturbed inner structure of the atmosphere. Finally, we note the presence of what appear to be instabilities at the boundary of the planetary wind and ambient medium at the sub-stellar regions of the flow, in contrast with the smooth flow in the low-flux runs.

The density panel in the $x-z$ plane (bottom left) again shows the significant oblique shock structure in the confined planetary wind torus, with the Mach surface located nearer the planet at the poles. The neutral density (bottom center) highlights both the greater extent of the neutral tail driven from the night side and the more extended recombination shadow of the planet. 

Figure \ref{fig:flow_texture_run3} shows the flow texture map for Run 3. Once again, the streamlines show the reorientation of the gas parcels launched from the planet's exobase due to Coriolis and tidal forces. Subtle differences in the flows between Run 3 and Run 1 can been seen in terms the orientation of the night-side shocks and neutral flow.

\subsection{Short-period planets: High Mass, High Flux Case}

The results for Run 4 are shown in figure \ref{fig:rxt_run4}. In the density panels, we note the strong similarities between the Mach surfaces for Run 4 and Run 2, with the two subsonic extensions in the up-orbit and down-orbit arms of the wind. Once again there is no high-density region extending beyond the night side to cause the ionization front to protrude beyond the planet's shadow. The neutral fraction panels display the truncation of the neutral tail, as seen in Run 2, and also show the recombination occurring in the shadow of the planet as in Run 3. In the neutral density in the $x-y$ plane, we see again the strong shock in the up-orbit arm that is absent in the down-orbit arm. In addition, we see the same shadow behind the planet due to recombination as in Run 3.

The density in the $x-z$ plane displays a new flow feature, wherein the subsonic area behind the planet is split into two lobes toward the north and south poles of the planet. In addition, we see as in figure \ref{fig:rxt_run3} the turbulence of the starward edge of the toroidally confined wind. The new lobe structure can also be seen in the bottom center panel, where the lobes are traced out by highly neutral material. Finally, the bottom right panel shows that the oblique shocks along the orbital direction have moved in significantly toward the planet on the night side.

Figure \ref{fig:flow_texture_run4} shows the flow texture map for the Run 4. The flow is generally similar to what is seen in the low flux, high mass case. Differences occur mainly in the details of where stagnation regions occur behind the planet and where shocks lead to redirection of the planetary wind.

\begin{figure*}
\centering
\includegraphics[width=\textwidth]{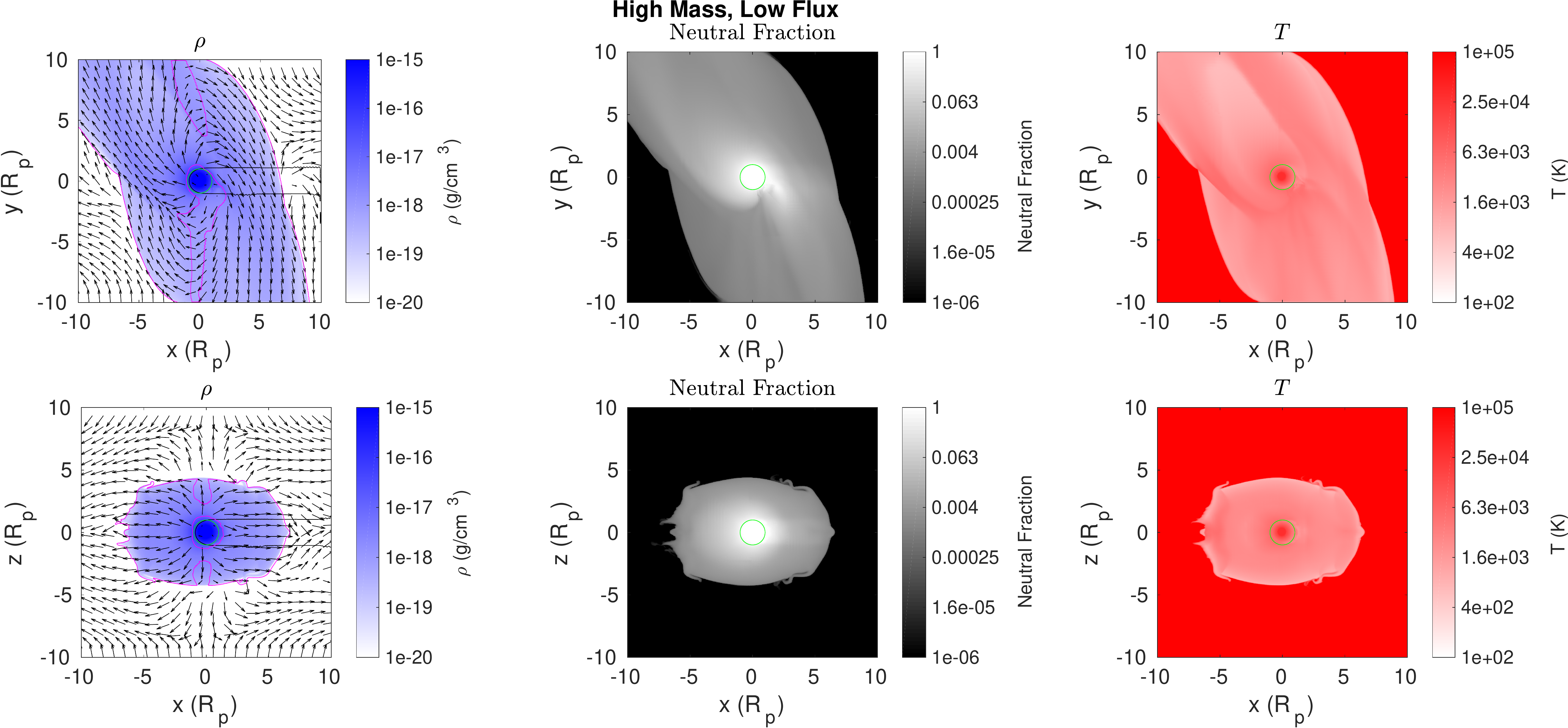}
\caption{Density (left), neutral fraction (center), and temperature (right) for Run 2 in the short period case, with a view down onto the orbital plane (top row) and standing in the orbital plane looking in the up-orbit direction (bottom row). The quivers describe the velocity field, and the contours are of the Mach surface (magenta), the $\tau = 1$ surface (black), and $R_p$ (green). Here the neutral tail is disrupted by material turned from the up-orbit arm by the Coriolis force, and the wind is more strongly confined along the radial and perpendicular directions.}
\label{fig:rxt_run2}
\end{figure*}

\begin{figure*}
\centering
\includegraphics[width=\textwidth]{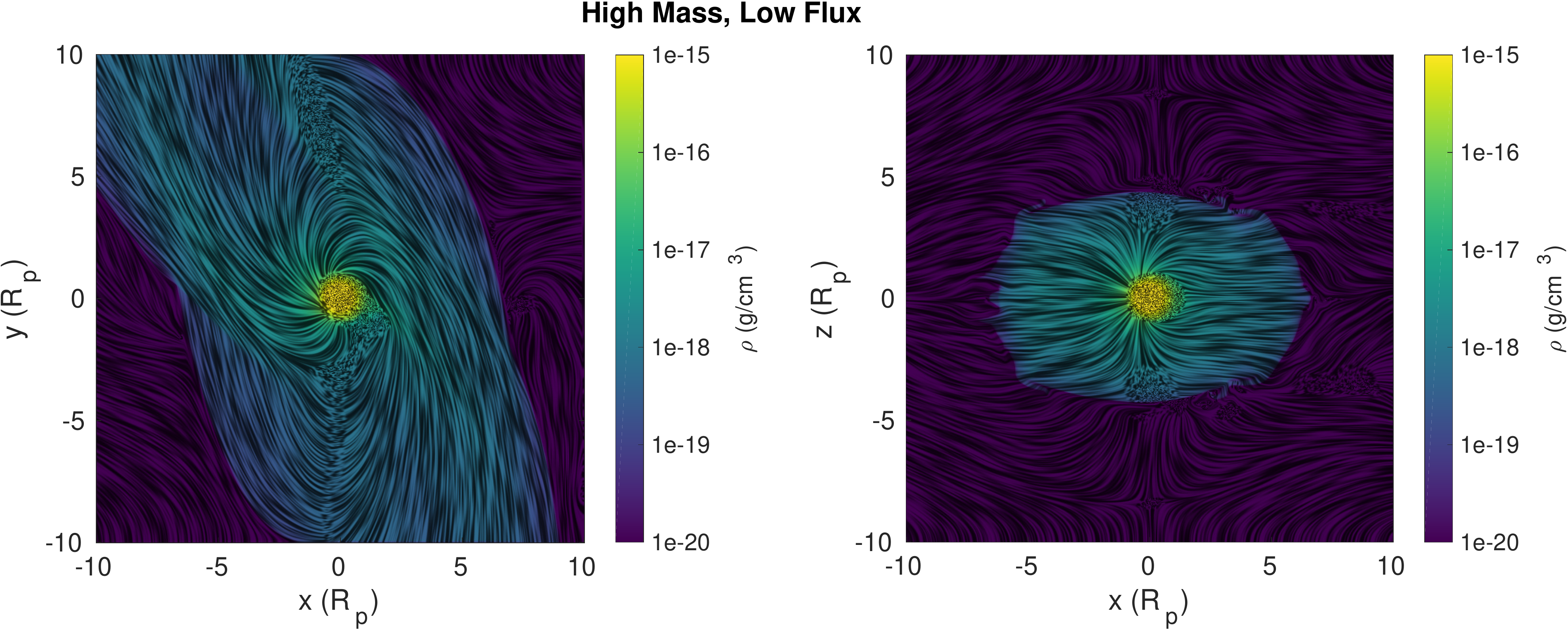}
\caption{Flow-texture plot of Run 2 in the short period case, showing density (hue) and local streamline orientation (texture), looking down onto the orbital plane (left) and standing in the orbital plane and looking up-orbit (right). The wind has a larger velocity here, and is therefore turned much more strongly than in Run 1. The azimuthal symmetry in the side view is limited due to this turning.}
\label{fig:flow_texture_run2}
\end{figure*}

\begin{figure*}
\centering
\includegraphics[width=\textwidth]{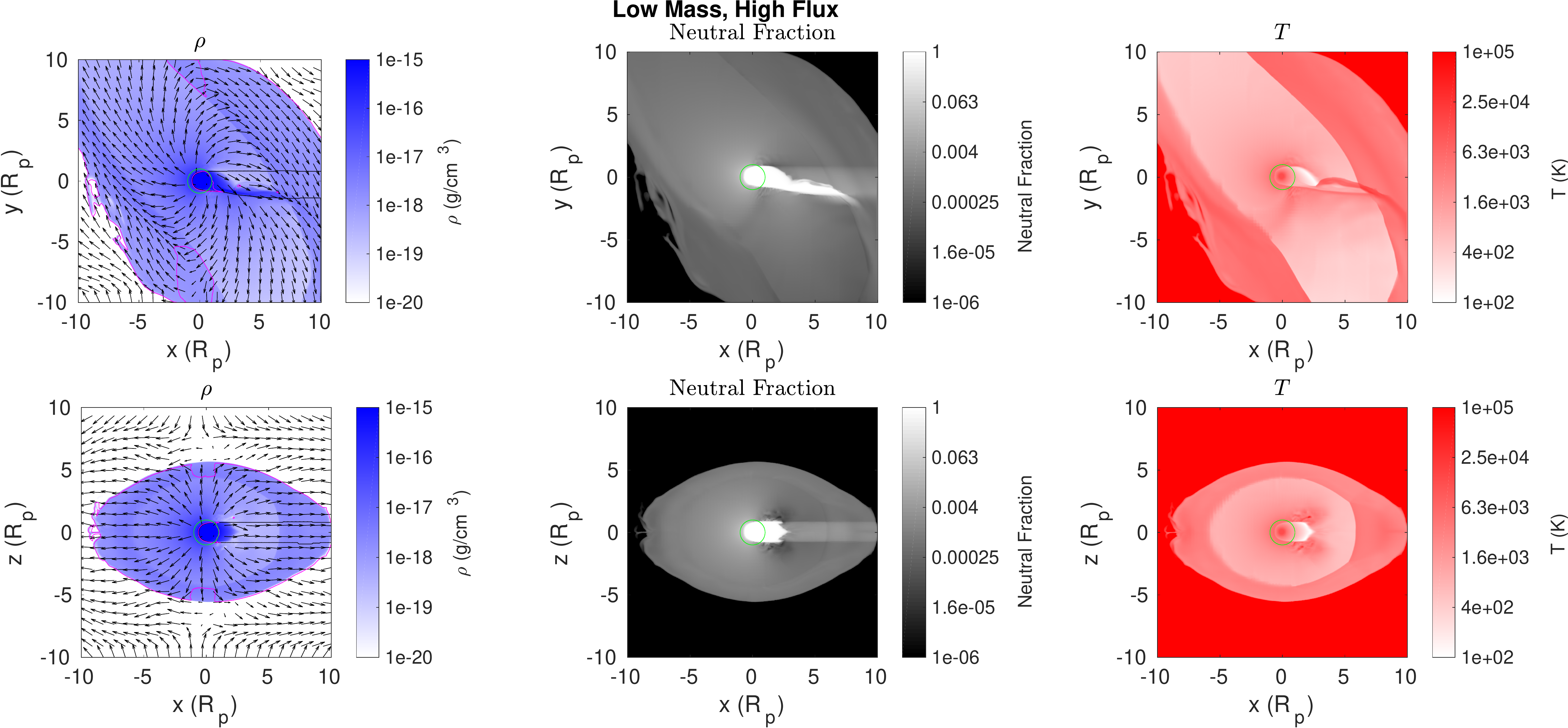}
\caption{Density (left), neutral fraction (center), and temperature (right) for Run 3 in the short period case, with a view down onto the orbital plane (top row) and standing in the orbital plane looking in the up-orbit direction (bottom row). The quivers describe the velocity field, and the contours are of the Mach surface (magenta), the $\tau = 1$ surface (black), and $R_p$ (green). Note again the presence of a dense tail of neutrals streaming from the night side of the planet, with greater cross-section than was found in Run 1.}
\label{fig:rxt_run3}
\end{figure*}

\begin{figure*}
\centering
\includegraphics[width=\textwidth]{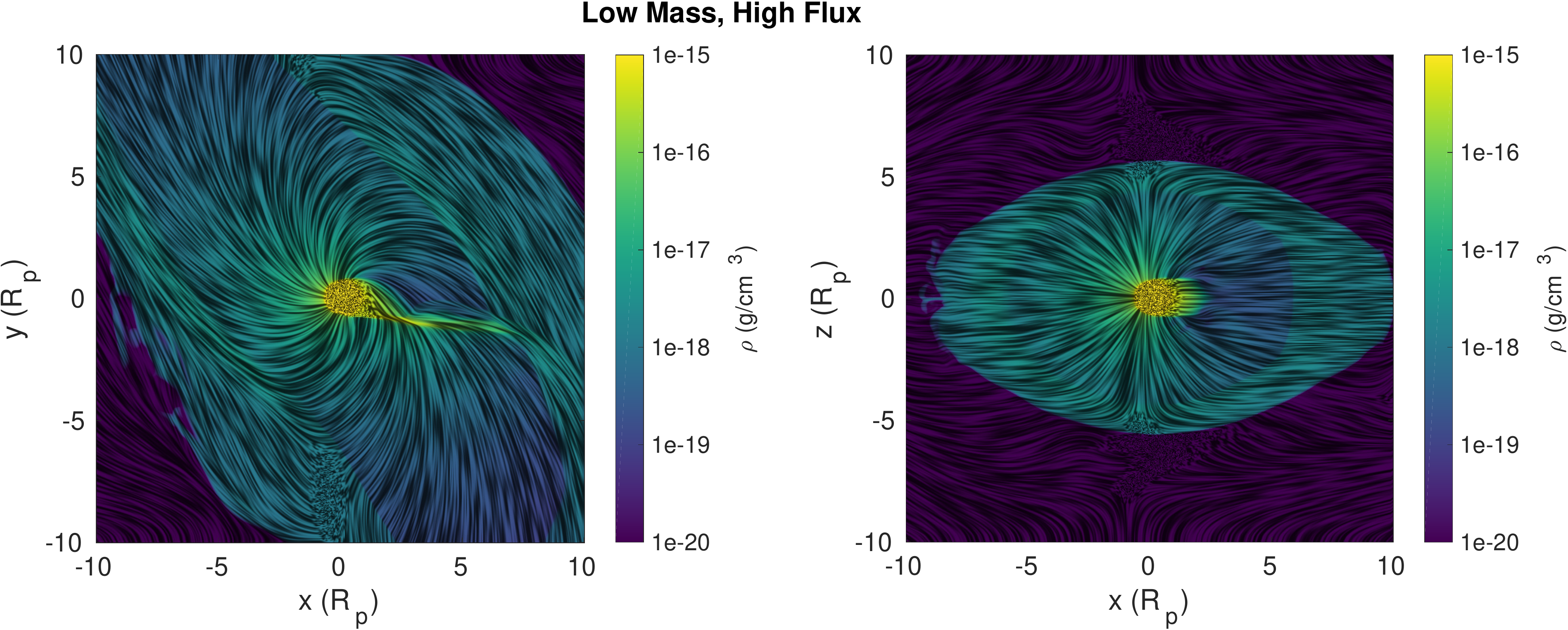}
\caption{Flow-texture plot of Run 3 in the short period case, showing density (hue) and local streamline orientation (texture), looking down onto the orbital plane (left) and standing in the orbital plane and looking up-orbit (right).}
\label{fig:flow_texture_run3}
\end{figure*}

\begin{figure*}
\centering
\includegraphics[width=\textwidth]{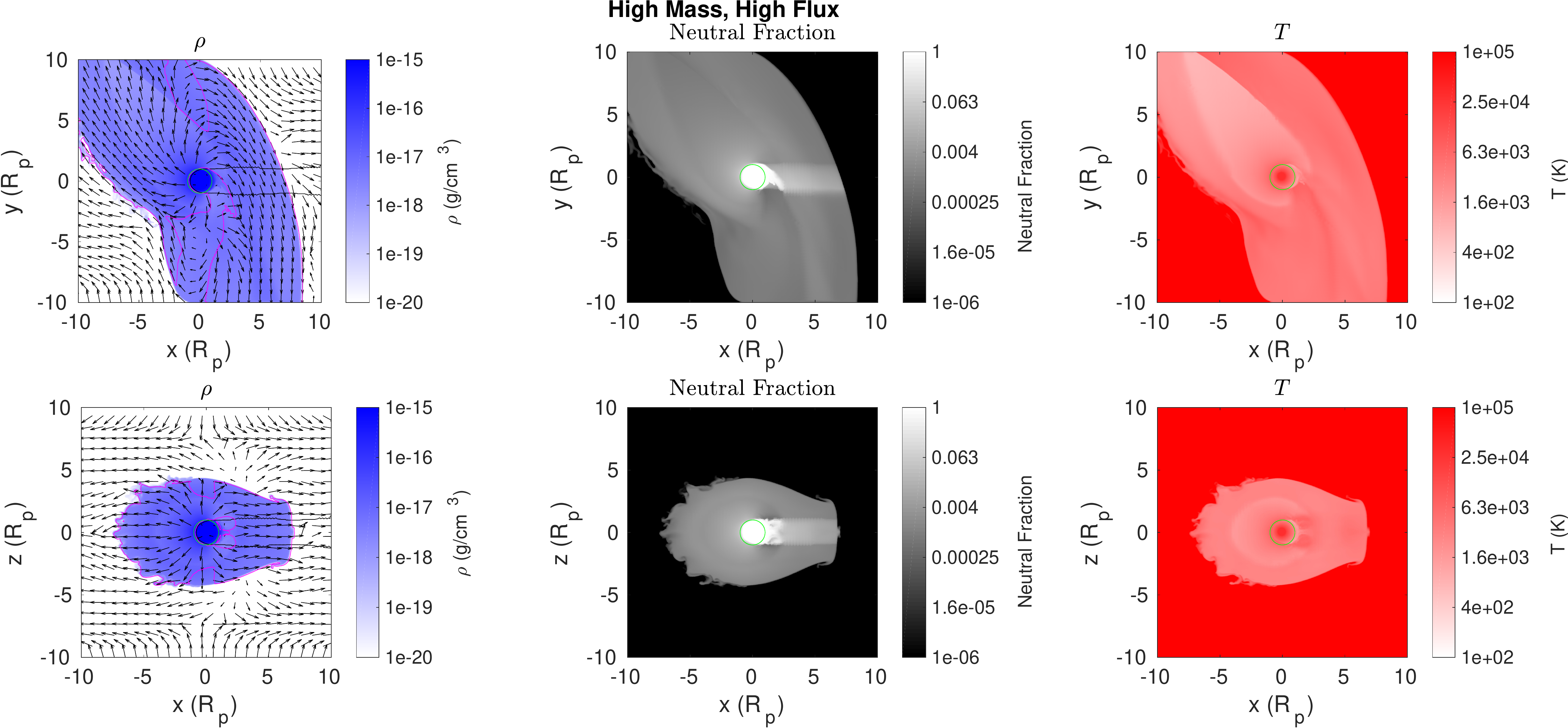}
\caption{Density (left), neutral fraction (center), and temperature (right) for Run 4 in the short period case, with a view down onto the orbital plane (top row) and standing in the orbital plane looking in the up-orbit direction (bottom row). The quivers describe the velocity field, and the contours are of the Mach surface (magenta), the $\tau = 1$ surface (black), and $R_p$ (green).}
\label{fig:rxt_run4}
\end{figure*}

\begin{figure*}
\centering
\includegraphics[width=\textwidth]{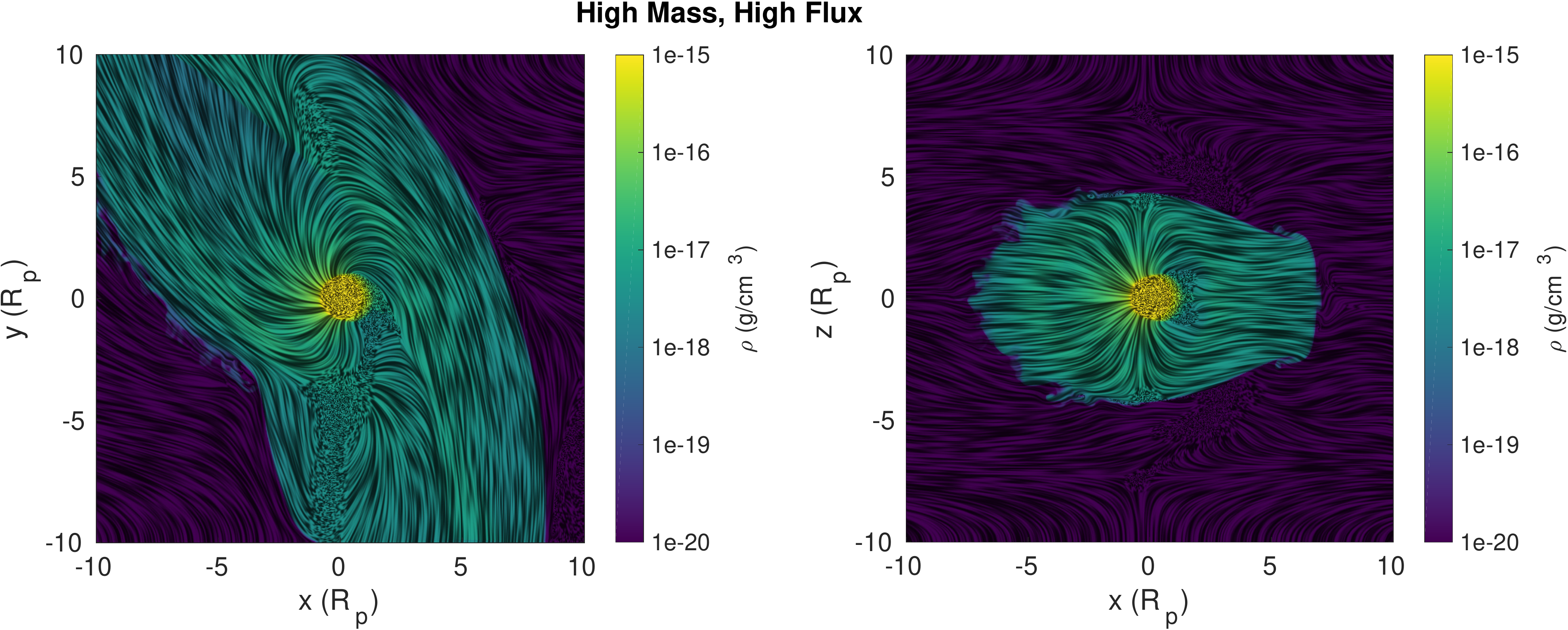}
\caption{Flow-texture plot of Run 4 in the short period case, showing density (hue) and local streamline orientation (texture), looking down onto the orbital plane (left) and standing in the orbital plane and looking up-orbit (right).}
\label{fig:flow_texture_run4}
\end{figure*}

\section{Analysis} \label{sec:ana}

\subsection{Wind Mass Loss Rates and Regimes} \label{sec:regime}

Photoevaporative winds can be classified based on the way incoming stellar flux is distributed in the flow \citep{murrayclay09,owen16}. For "energy-limited" mass loss, most of the energy deposited by photoionization as heat is $\propto F_{UV} R_p^2$ and goes into $PdV$ work. Losses due to radiation and internal energy changes are small. The $PdV$ work, measured per unit mass, can be expressed as
\begin{equation}
 \frac{P \delta V}{\rho R_p^2 H} \sim \frac{G M_p}{R_p},
\end{equation}
which yields an energy-limited mass loss rate of 
\begin{equation}
\hspace*{-25pt}
 \dot{M}_{el} \sim 6 \times 10^9 \left(\frac{\epsilon}{0.3}\right)
 \left(\frac{R_p}{10^{10}cm}\right)^3\left(\frac{0.7 M_J}{M_p}\right)
 \left(\frac{F_{UV}}{450 \mbox{ erg cm}^{-2}}\right) \mbox{ g s}^{-1}.
\end{equation}
Note that the mass loss rate is linear with $F_{UV}$. In the more detailed numerical models of \citet{murrayclay09} the actual dependence was slightly weaker: $\dot{M}_{el} \sim F_{UV}^{0.9}$.

\begin{figure}
\centering
\includegraphics[width=\columnwidth]{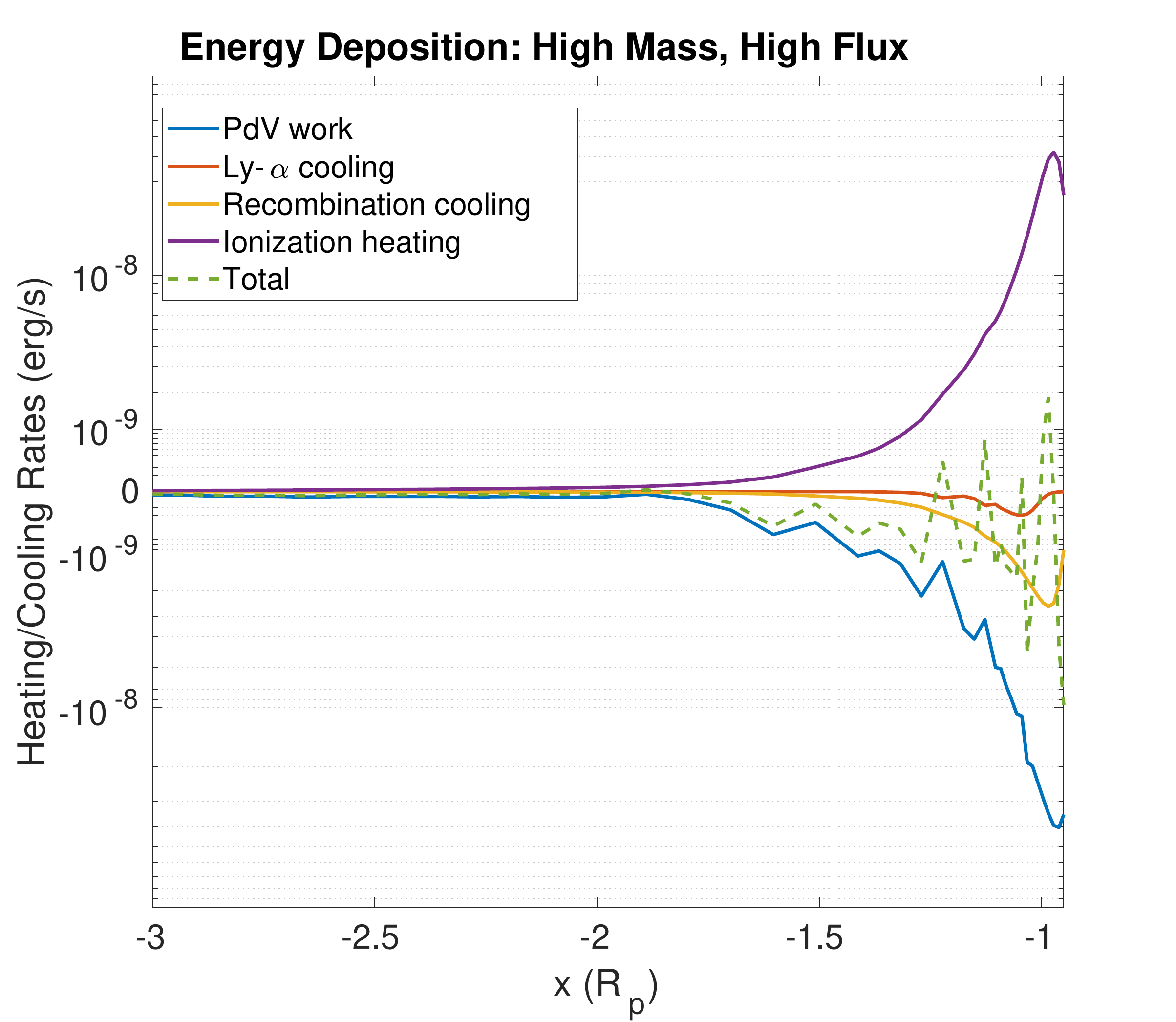}
\caption{Plot of the heating and cooling terms along the star-planet axis for Run 4. Because the cooling is primarily by expansion (PdV work), we see that we are in the energy-limited regime. Also note the compressive heating on the night side of the planet.}
\label{fig:energy_balance_run4}
\end{figure}

Alternatively, for "radiation/recombination limited" mass loss, the photoevaporative flows are such that the input UV power is mostly lost to radiative cooling. The balance of radiative heating and cooling then keeps the gas temperature at $T \sim 10^4$ K. In this case it can be shown that the mass loss rate goes as 
\begin{equation}
 \dot{M}_{rl} \sim 4 \times 10^{12} 
 \left( \frac{F_{UV}}{5 \times 10^5 \mbox{ erg cm}^{-2}}\right)^{1/2} \mbox{ g s}^{-1}.
\end{equation}
We can examine this question directly with figure \ref{fig:energy_balance_run4}, which shows the relative size of heating and cooling terms in one of our runs. The remainder are nearly identical. From this plot it is clear that all of the cases examined in this study are in the energy-limited regime, with nearly all of the energy deposited by EUV radiation being used to launch the wind.

Mass loss rates for each run were calculated by integrating the flux through a spherical surface at $3 R_p$. Despite the significant differences in flow structure, the mass loss rates of each low flux run were similar. We find mass loss rates of between $3\times10^{10}$ and $4\times10^{10}$ for the low flux cases and approximately $2\times10^{11}$ for the high flux cases, comparable to the mass loss rates found in \citet{tripathi15} for the same radiation flux. Note that we see the wind mass loss increasing with flux at a lower rate than predicted for an energy-limited flow $\dot{M} \sim F_{UV}^{.9}$, which for a given factor of 10 increase should yield a mass loss increase factor of 7.9. In the models we find ${\dot M}_{Run3}/{\dot M}_{Run1} \sim 4.39$ and ${\dot M}_{Run4}/{\dot M}_{Run2} \sim 6.09$. However, because the mass loss rate models were derived with a planet of constant radius, the fact that our winds were launched from slightly different radii in each run explains the difference in mass loss rates here. If we include the radius dependence, $\dot{M} \sim R_{\tau=1}^3 F_{UV}^{.9}$ \citep{murrayclay09}, we find theoretical mass loss ratios of ${\dot M}_{Run3}/{\dot M}_{Run1} \sim 3.53$ and ${\dot M}_{Run4}/{\dot M}_{Run2} \sim 6.39$, similar to our measured values.

\begin{table}
 \begin{minipage}{80mm}

 \caption{Mass loss rates}
 \label{tab:mdot}
 \begin{tabular}{c|c|}

  Simulation & $\dot{M}$ \\
  & $\mbox{g s}^{-1}$ \\

  \hline

  Low Mass, Low Flux, non-rotating & $3.42 \times 10^{10}$ \\
  High Mass, Low Flux, non-rotating & $3.11 \times 10^{10}$ \\
  Low Mass, Low Flux & $3.87 \times 10^{10}$ \\
  High Mass, Low Flux & $3.35 \times 10^{10}$ \\
  Low Mass, High Flux & $1.70 \times 10^{11}$ \\
  High Mass, High Flux & $2.04 \times 10^{11}$ \\

  \hline

 \end{tabular}
 \end{minipage}
\end{table}

\subsection{Rotational effects} \label{sec:rotation}

In order to discuss the effects of rotation on the wind, we define a few quantities (following \citet{carroll16}). $\phi_c$ is the critical angle (measured clockwise from the substellar ray) for determining whether material leaving the planet joins the up-orbit or down-orbit arm, given by
\begin{equation}
 \phi_c = \arctan\left(\frac{-2 r \Omega}{v}\right).
\end{equation}
For our analysis, we take the radius and sound speed at the sonic radius along the substellar ray, which leads to a slight overestimate of the value of $\phi_c$. In addition, the orbital angle of the subsonic eddy which occurs in the up-orbit arm of the wind is given by
\begin{equation}
 \Theta_D = \frac{12 v}{\Omega a},
\end{equation}
where the speed is again taken at the sonic radius along the substellar ray.
These quantities are summarized for the suite of simulations in Table \ref{tab:rot}.

\begin{table}
 \begin{minipage}{80mm}

 \caption{Rotational quantities}
 \label{tab:rot}
 \begin{tabular}{c|c|c}

  Simulation & $\phi_c$ & $\Theta_D$ \\

  \hline

  Low Mass, Low Flux & $-0.25 \pi$ & $0.139 \pi$ \\
  High Mass, Low Flux & $-0.24 \pi$ & $0.173 \pi$ \\
  Low Mass, High Flux & $-0.17 \pi$ & $0.156 \pi$ \\
  High Mass, High Flux & $-0.19 \pi$ & $0.198 \pi$ \\

  \hline

 \end{tabular}
 \end{minipage}
\end{table}

It is noteworthy that our simulations find the same rotational effects as found in the \textit{Rotating} case of \citet{mccann18}, conducted using the hydrodynamics code Athena. In particular, note the similarities between figures 2(c) and (f) and our figure \ref{fig:rxt_run2}.

\subsection{Planetary ionization shadow}

\begin{figure}
    \centering
    \includegraphics[width=\columnwidth]{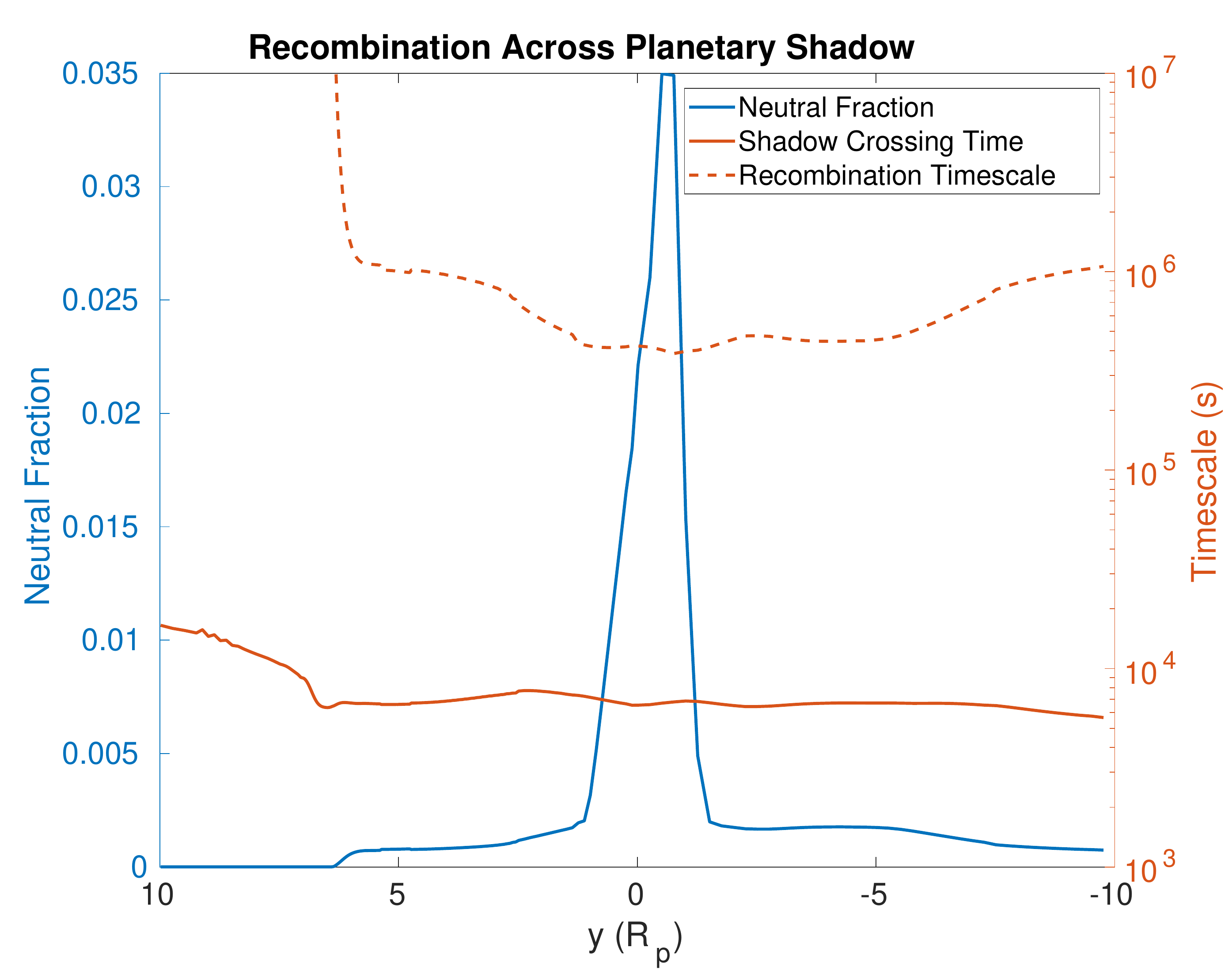}
    \caption{Neutral fraction (left axis) and recombination and shadow crossing timescales (right axis) for Run 4, taken at $x=4 R_p$. As we enter the planetary wind material, the recombination timescale drops sharply and the neutral fraction increases to its steady-state bulk wind value of approximately 0.002. Upon entering the planet's shadow around $1 R_p$, the neutral fraction begins to rise as the ionizing radiation is blocked. Exiting the planet's shadow around $-1 R_p$, the wind material ionizes precipitously as it is again exposed to the ionizing radiation.}
    \label{fig:recom_shadow}
\end{figure}

The change in the neutral fraction behind the planet seen in Runs 3 and 4 can be understood in terms of the shadowing of the higher flux by the planet. The shadow is enhanced by the recombination of material temporarily protected by the planet from ionizing radiation. We can compare the recombination timescale $\tau_{recom}$ and the shadow crossing time $\tau_{shadow}$ for Run 4 to get an estimate of the expected neutral fraction on the far side of the tail:
\begin{equation}
 \tau_{recom} = \frac1{\alpha_B n_e} = 4.2 \times 10^5 \mbox{ s},
\end{equation}
\begin{equation}
 \tau_{shadow} = \frac{2 R_p}{v_y} = 1.34 \times 10^4 \mbox{ s}.
\end{equation}
This gives us an estimate of the total change in neutral fraction over the length of the shadow of $\tau_{shadow}/\tau_{recom} = 0.032$. As figure \ref{fig:recom_shadow} demonstrates, the neutral fraction climbs to 0.035 at the edge of the shadow before dropping sharply as expected.

It is also possible to determine whether there is a "sweet spot" in parameter space such that the recombined material from the planet's shadow remains neutral for a sufficiently long time to be detectable after it passes out of the shadow (in the absence of a denser neutral tail, such as in Runs 2 and 4). Using the ionization timescale,
\begin{equation}
    \tau_{ion} = \frac1{\sigma_{ph} F_0} = 800 \mbox{ s},
\end{equation}
we can calculate the distance we would expect it to take to return to the equilibrium wind ionization:
\begin{equation}
    x = \tau_{ion} v_y = 1.8 \times 10^9 \mbox{ cm} = 0.12 R_p.
\end{equation}
In fact, it takes $\sim \frac34 R_p$ for the material to ionize back to the equilibrium level of the wind in Run 4. The extended distance here can be partially explained by the fact that the wind is still recombining as it exits the shadow, so that the ionization fraction does not increase as quickly as it would without the recombination. Therefore, it may be possible to detect the material that has recombined in the planet's shadow even for flux as high as we use here.

\subsection{Neutral tail}

\begin{figure}
    \centering
    \includegraphics[width=\columnwidth]{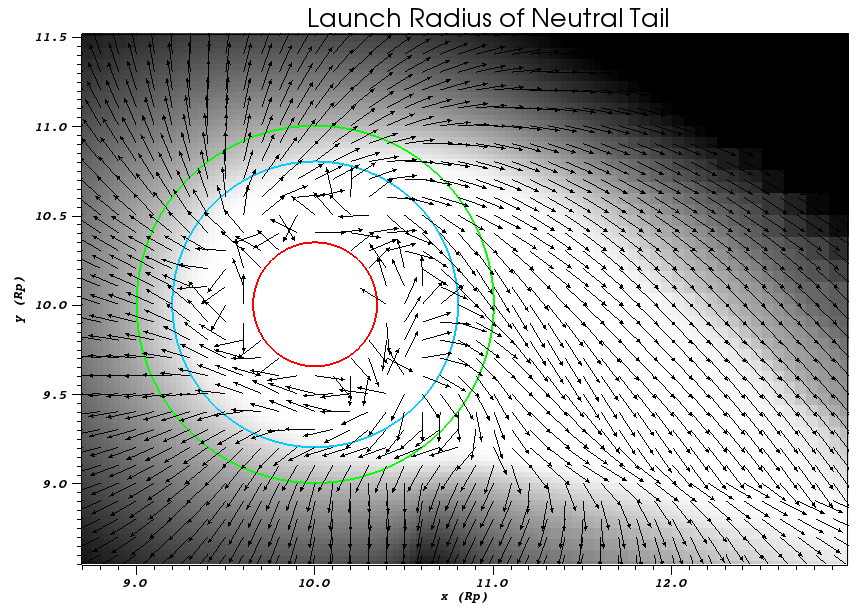}
    \caption{A close-up of the steady state of the planet in Run 1, with the red contour showing $R_{ib}$, the radius held fixed during the simulation, the blue contour showing the radius of the ionization front at the substellar point $R_{\tau=1}=0.8 R_p$, and the green contour showing the nominal planet radius $R_p$. The vectors show the velocity field, and the hue is the neutral fraction. Note that much of the material in the neutral tail appears to be originating within the blue contour.}
    \label{fig:tail_launch}
\end{figure}

In section \ref{sec:result}, we identified the importance of neutral material streaming laterally from the night side of the planet, as well as its absence in the simulations with high planet mass (Runs 2 and 4). While we don't believe any such phenomenon has yet been detected, it could provide useful diagnostic information about the planet. The vector field of velocities in figure \ref{fig:tail_launch} provide a guide to the depth at which the neutral tail material originates. In addition to its presence as an indicator of low planet mass, the fact that the material in the tail is drawn from within the planet could allow its use as a probe into the composition of the planet at deeper radii than we are currently able to detect.

\section{Discussion and Conclusions} \label{sec:conc}

In the preceding analysis, we have studied the effects of planet mass and stellar ionizing flux on photoevaporative winds launched from gaseous giant planets using self-consistent radiative transfer. We found that low mass planets produce a potentially observable extended neutral tail of material being advected from within the night side of the planet that is not present in the simulations of high mass planets. We also see that low mass planets create a torus of wind material of larger cross section than that produced by high mass planets. In addition, the high flux simulations result in a more pronounced recombination region in the ionization shadow of the planet than seen in the low flux cases. Finally, we observed mass loss rates a factor of approximately 5x greater in the high flux cases than those seen in the low flux simulations.

Here we compare to two previous works which investigated the physics most relevant to the current study, \citet{carroll16} and \citet{tripathi15}. We begin with \citet{carroll16}. In that study the planetary wind was launched from a fixed pressure and density boundary condition rather than by deposition of energy via ionizing radiation. We have extended that study by including a wind launched by self-consistent radiative heating in both rotating and non-rotating frames. We first compare panel ANISO of figure 3 of \citet{carroll16} (which shows the effect of orbital motion on wind streamlines) to the flow-texture plots of our simulations. The comparison shows that the streamlines close to the planet match well, particularly in the low mass planet case.

We also wish to investigate whether orbital motion (i.e., the Coriolis and centrifugal forces) has any effect on the mass loss rate from the planet. In \citet{carroll16}, it was found that uneven planetary heating (intrinsic in tidally-locked planets) reduced the mass loss rate by almost a factor of 2, but that placing an unevenly-heated planet into a rotating frame of reference had no effect on the mass loss rate. Although we did not run a simulation with uniform planetary heating, we also find that rotation has little effect on the mass loss rate through the wind, as the mass loss rates for Runs 1 and 2 in the rotating and non-rotating cases differ by only 10\% (see table \ref{tab:mdot}).

We now compare the results of our study that of \citet{tripathi15}, who performed simulations of the radiative heating of the atmosphere of similar planets in the presence of tidal forces but without non-inertial forces, which we have now included. In order to compare similar simulations, we consider the long-period planet with high planet mass, with no rotational or tidal effects (figure \ref{fig:rxt_run1norot}. Note that our masses differ by almost a factor of two and our surface densities by an order of magnitude. \citet{tripathi15} found that the sonic surface was significantly farther from the planet, at a distance of approximately $0.6 R_p$ compared to a distance of $0.1 R_p$ for our simulation (compare \citet{tripathi15} figure 3 and the left panel of figure \ref{fig:rxt_run2norot}). This difference is likely explained by the lower planet mass, which we have seen in comparing our own models leads to a decrease in the sonic radius. We can also compare the overall flow patterns in the wind, shown in the bottom left panel of figure 2 in \citet{tripathi15} and figure \ref{fig:flow_texture_run2norot}. Although the flow patterns are broadly similar, with axisymmetric outflow and lateral flows toward the night side, the lateral flows are flowing back slightly more strongly onto the planet in the previous study.

We find a mass loss rate approximately 1.4x larger for the low flux case than \citet{tripathi15}. Although we did not simulate the high flux cases without rotational effects, as argued above, the addition of rotation has little effect on the mass loss rate and we can conclude that the high flux non-rotating case would display a mass loss rate of approximately $2\times10^{11}$ g/s, nearly identical to that found in the high-flux case of \citet{tripathi15}.

Finally, we can also locate our simulated planetary winds in the regimes identified by both \citet{murrayclay09} and \citet{matsakos15}. As discussed in section \ref{sec:ana}, all of our simulations are located in the energy-limited regime rather than the radiation/recombination (or cooling) limited regime, with the majority of the input EUV energy being used for the expansion of the wind rather than radiative cooling. In addition, we have no stellar wind or magnetic fields. Our low ambient pressure and relatively low planet mass (even in the high mass cases) allow us to place all of our simulations in the "Type III" regime of \citet{matsakos15}, with the wind flow dominating planetary tidal forces. To produce their Type I or Type IV interactions, a strong magnetic field and/or a weak planetary wind (relative to the ambient/stellar wind pressure) are required. This confines the wind, preventing it from forming the up-orbit arm. A high-mass planet might produce a Type I interaction, with the wind confined to the vicinity of the planet, while a low-mass planet would produce a Type IV interaction, with the wind overflowing the Roche lobe and being captured by the star. To produce a Type II interaction, a higher-mass planet or stronger stellar wind pressure would be required, so that the wind material is again confined to the vicinity of the planet rather than being captured by the star. The fact that our interactions are Type III means we would expect in a simulation of the full system a long up-orbit arm extending a significant distance around the star with accretion at the far end of this arm, as well as an extended tail.

\subsection{Phenomena not yet considered}

There are a number of physical processes which we have not considered in these simulations, but which may have important consequences for the structure of the wind and for the observational signatures of evaporating planets. The interaction of the planetary wind with a strong stellar wind has been shown to be able to confine the up-orbit arm of the wind \citep{matsakos15, Schneiter2017, mccann18}. Radiation pressure has also been shown to be a potentially important mechanism for the acceleration of neutral hydrogen to speeds required to reproduce the observed Lyman-$\alpha$ absorption \citep{Schneiter2017,bourrier15}. Another mechanism proposed to produce such a fast neutral population is charge exchange between the ionized stellar wind and the planetary wind \citep{bourrier16,tremblin13,christie16}. Magnetic fields are also expected to affect both the launching of the wind and its interaction with the stellar wind \citep{owen14,matsakos15,villarreal18}. Finally, we note that it has recently been proposed that the metastable helium line at 10830 \AA will provide an excellent source of observational diagnostics for evaporating exoplanets, since it is relatively abundant and unaffected by ISM absorption \citep{oklopcic18}. This requires modeling the photoionization dynamics of additional species. 

Only \citet{mccann18} of the works above, however, has included a self-consistent treatment of photo-ionization in driving the wind from the planet's atmosphere as was done in this study. Thus many of the listed processes can be added productively to the kind of simulation platform developed here.

Although it is a more difficult computational problem, it will also be important to include the effects of spherical dilution of the radiation field in order to examine the full orbital behavior of the wind material. As an example of a phenomenon where the full orbital treatment is required, accretion onto the star (as seen in \citet{matsakos15}) could provide another important observational diagnostic for the behavior of evaporating hot Jupiters. In addition, the full orbital treatment is likely to be important for cases such as WASP-12b, where we have previously shown \citep{debrecht18} that it is possible for a torus to form completely surrounding the star. Such a torus is likely to be significantly affected by radiation pressure.

\section{Acknowledgments}

We thank Luca Fossati for many helpful conversations. We also thank the Other Worlds Laboratory (OWL) at University of California, Santa Cruz for facilitating this collaboration by way of the OWL Exoplanets Summer Program, funded by the Heising-Simons Foundation. This work used the computational and visualization resources in the Center for Integrated Research Computing (CIRC) at the University of Rochester and the computational resources of the Texas Advanced Computing Center (TACC) at The University of Texas at Austin, provided through allocation TG-AST120060 from the Extreme Science and Engineering Discovery Environment (XSEDE) \citep{xsede}, which is supported by National Science Foundation grant number ACI-1548562. Financial support for this project was provided by the Department of Energy grant DE-SC0001063, the National Science Foundation grants AST-1515648 and AST-1411536, and the Space Telescope Science Institute grant HST-AR-12832.01-A.

\bibliography{planets.bib}

\bsp

\label{lastpage}

\end{document}